\documentclass[twocolumn,aps,superscriptaddress,multicol,amsmath,amssymb]{revtex4-2}
\usepackage{mathtools}
\usepackage{mathtools}
\usepackage{bm}
\usepackage{comment}
\usepackage{dsfont,amsthm,amsbsy}
\usepackage{verbatim}
\usepackage{amssymb}
\usepackage{amsmath}
\usepackage{bbm}
\usepackage{graphicx}
\usepackage{epstopdf}
\usepackage{subfigure}
\usepackage{natbib}
\usepackage{epsfig}
\usepackage{amsfonts}
\usepackage{mathrsfs}
\usepackage{sidecap}
\usepackage{lipsum}
\usepackage[toc,page,title,titletoc,header]{appendix}
\usepackage[colorlinks,linkcolor=blue,citecolor=blue,anchorcolor=blue,urlcolor=blue]{hyperref}
\usepackage{hyperref}
\usepackage{resizegather}
\usepackage{tikz}
\usepackage{float}
\usepackage{mathbbol}
\usepackage[normalem]{ulem}
\usepackage{cancel}
\usepackage{upgreek}

\newcommand{\bl}{\begin{aligned}}
\newcommand{\el}{\end{aligned}}
\def\be{\begin{equation}}
\def\ee{\end{equation}}

\def\bi{\begin{itemize}}
\def\ei{\end{itemize}}
\def\bn{\begin{enumerate}}
\def\en{\end{enumerate}}
\def\bea{\begin{eqnarray}}
\def\eea{\end{eqnarray}}
\def\no{\nonumber}
\def\ba{\begin{array}}
\def\ea{\end{array}}
\def\bd{\begin{displaymath}}
\def\ed{\end{displaymath}}

\graphicspath{{./}{./Figs/}}

\begin{document}

\title{Dynamical Quantum Phase Transitions in a Pseudo-Hermitian Hamiltonian: The Imbalanced-Pairing Kitaev Model}

\author{R. Jafari}
\email[]{raadmehr.jafari@gmail.com}
\affiliation{Department of Physics, Institute for Advanced Studies in Basic Sciences (IASBS), Zanjan 45137-66731, Iran}
\affiliation{School of Quantum Physics and Matter Science, Institute for Research in Fundamental Sciences (IPM), 19395-5531, Tehran, Iran}

\author{Alireza Akbari}
\email[]{alireza@bimsa.cn}
\affiliation{Beijing Institute of Mathematical Sciences and Applications (BIMSA), Huairou District, Beijing 101408, China}
\affiliation{Max Planck Institute for the Chemical Physics of Solids, D-01187 Dresden, Germany
}

\author{Shukhrat Mardonov}
\email[]{sh.mardonov@newuu.uz}
\affiliation{New Uzbekistan University, Movarounnahr Street 1, Tashkent 100000, Uzbekistan}

\author{H. Yavartanoo}
\email[]{yavar@bimsa.cn}
\affiliation{Beijing Institute of Mathematical Sciences and Applications (BIMSA), Huairou District, Beijing 101408, China}

\begin{abstract}
Although parity-time (PT)-symmetric Hamiltonians are often associated with real energy spectra, PT symmetry is neither a sufficient nor a necessary condition for a real spectrum. More generally, real spectra are associated with the broader class of pseudo-Hermitian Hamiltonians, of which PT-symmetric Hamiltonians constitute a simple subclass.
Here, we investigate the nonequilibrium dynamics of the imbalanced-pairing Kitaev model, a prototypical pseudo-Hermitian system, under a linearly time-dependent chemical potential. The dynamics are analyzed within the biorthogonal framework using the concept of dynamical quantum phase transitions (DQPTs).
We show that, under a linear ramp protocol, DQPTs occur only when the post-ramp Hamiltonian possesses a real energy spectrum.
For positive values of the non-Hermiticity parameter ($\gamma>0$), where the energy spectrum remains entirely real, a ramp crossing a single quantum critical point gives rise to a single family of critical times, analogous to the Hermitian case.
Furthermore, for ramps crossing two critical or exceptional points, the
critical sweep velocity above which DQPTs disappear decreases as the
non-Hermiticity parameter is reduced and vanishes in the staggered-pairing
limit, $\gamma=-1$.
\end{abstract}

\maketitle

\section{Introduction}

Recent advances in experimental techniques have enabled the investigation of nonequilibrium quantum dynamics in a highly controlled manner~\cite{Dupont2022,Langen2015,king2022,keesling2019}. Consequently, nonequilibrium quantum phenomena have attracted considerable interest in both theoretical and experimental studies over the past decades~\cite{Polkovnikov2011,Jafari2017}. This progress has led to the discovery of a wide range of intriguing phenomena, including the Kibble-Zurek mechanism~\cite{Dziarmaga2005,Francuz2016}, discrete time crystals~\cite{Yang2021}, many-body localization~\cite{Abanin2019}, ergodicity breaking~\cite{Mishra2013,Chanda2016,Awasthi2018}, and dynamical quantum phase transitions (DQPTs)~\cite{Heyl2013,Heyl2017,Heyl2018,Bhattacharya}.

DQPTs have emerged as one of the central themes in nonequilibrium many-body physics and have been extensively studied in recent years~\cite{Jafari2019,Sadrzadeh2021,Wong2022,Corps2022,Corps2023,Francisco2022,Bhattacharjee2018,Bhattacharya2017,Xu2026,Parez2026a,Parez2026b,Novotny2026,Xu2023,Luo2026,Zhang2026,Domagoj2025,Celeri2026,Kheiri2026,Liu2026QB,Bento2024,Goes2020,Bozhen2019,Ye2025}. Since their introduction in the Hermitian transverse-field Ising model~\cite{Heyl2013}, DQPTs have been generalized to mixed states~\cite{Heyl2018,Bhattacharya}, finite temperatures~\cite{Abeling2016,Lang2018a,Lang2918b,Mera2018,Tang2025}, Floquet systems~\cite{Yang2019,Zamani2020,Naji2022,Jafari2021,Kosior2018a,Jafari2022,Kosior2018b,Naji2022b,Cheng2026}, sudden quenches~\cite{Sedlmayr2023,Sedlmayr2022,Sedlmayr2020,Uhrich2020,Zeng2023,Stumper2022,Yu2021,Vijayan2023,Xue2023,Bhattacharjee2023,Leela2022,Puskarov2016,Haldar2020,Ye2024,Jafari2019b,Mitra2026,Bozhen2021,Cao2024,Tian2020,Schmitt2015}, ramp protocols~\cite{Divakaran2016,Zamani2024,Zheng2026,Jafari2024,Baghran2024,Ansari2025,Jafari2025arxiv,Kheiri2025,Zhang2025,XiangZhang2026,Chen2020}, disordered systems~\cite{Mishra2020,Vanhala2023,Cao2020}, and dynamics under correlated~\cite{Jafari2024} and uncorrelated noise~\cite{Baghran2024,Ansari2025,Kheiri2025,Jafari2025arxiv}. They have also been experimentally observed in several platforms, including trapped ions~\cite{Jurcevic2017,Zhang2017,Martinez2016,Neyenhuis2017,Smith2016}, Rydberg atoms~\cite{Bernien2017}, ultracold atoms~\cite{Flaschner2018}, superconducting qubits~\cite{Guo2019}, and nanomechanical and photonic systems~\cite{Wang2019,Xu2020}, nuclear magnetic resonance quantum simulator~\cite{Nie2020}.

DQPTs are associated with nonanalyticities in the dynamical free-energy density, with real time playing the role of the control parameter~\cite{Vajna2015,Karrasch2013,Vajna2014,Mendoza2022,Tomasz2024,Maslowski2026,Bhattacharyya2026,Sedlmayr2019,Khatun2019,Ding2020,Nicola2021,Verga2023,Rossi2022,Khan2023,Khan2023,Jad2021,Vajna2014,Porta2020,Wang2018,Yang2026}. These nonanalyticities arise when the Loschmidt amplitude vanishes in the complex-time plane, corresponding to the time-evolved state becoming orthogonal to the initial state. In addition, DQPTs can be characterized by jumps in a winding number, which serves as a dynamical topological marker of the real-time evolution~\cite{Budich2016,Lang2018a}.

Dissipation and environmental coupling provide another important route to nonequilibrium quantum phenomena, motivating extensive studies of open quantum systems and dissipative phase transitions~\cite{Prosen2008,Dagvadorj2015,Jin2016,Bakker2022,Kazmina2024}. Recent experimental progress has made it possible to engineer and control dissipation in platforms such as ultracold atoms~\cite{Berman2012}, trapped ions~\cite{Leibfried2003}, quantum optical systems~\cite{Miri2019}, and superconducting circuits~\cite{Clerk2020}. These developments have opened new directions for exploring dissipative and non-Hermitian quantum systems, with possible applications in quantum technologies~\cite{Harrington2022}.

Non-Hermitian systems can be realized in a variety of experimental platforms, ranging from photonics~\cite{Xiao2020} and phononics~\cite{Zhu2018} to nitrogen-vacancy centers~\cite{Yang2019} in solids and cold atoms~\cite{Gou2020}. These platforms have enabled the exploration of phenomena without direct Hermitian counterparts, including real spectra in non-Hermitian Hamiltonians with parity-time (PT) symmetry~\cite{Bender1998}, the non-Hermitian skin effect~\cite{Yao2018,Xiujuan2022}, and complex non-Hermitian topology~\cite{Bergholtz2021,Ashida2020,Shen2018,Kawabata2019,Okuma2023,Zhang2026DM}. In quantum phase transitions (QPTs), non-Hermiticity introduces new spectral critical phenomena associated with PT symmetry and intrinsic rotation-time-reversal (RT) symmetry~\cite{XZZhang2013,CLi2014,TonyLee2014}. When one of these symmetries is preserved, the spectrum can remain purely real, whereas symmetry breaking can drive it into a complex-spectrum regime. The transition is associated with exceptional points, where the eigenvalues become degenerate and the corresponding eigenvectors coalesce~\cite{Heiss2012}. Such exceptional points play a central role in non-Hermitian physics and have been linked to enhanced sensitivity in several settings~\cite{Miri2019}.

Dynamics generated by non-Hermitian Hamiltonians exhibit a rich variety of phenomena and have attracted significant theoretical and experimental attention~\cite{Wang2019,KLZhang2021}. DQPTs in non-Hermitian systems have been studied using self-normalized states and related normalization procedures~\cite{Zhou2018,Zhou2021,Das2024,Mondal2024,Mondal2022,Tang2026}. More recently, the biorthogonal framework has been applied to non-Hermitian dynamics~\cite{Jing2024,Wang2026b,Gu2026,Fu2025,Deng2025}, where the biorthogonal Loschmidt echo is naturally normalized. 
Most previous studies have focused on sudden quenches and
particle-number-conserving single-particle models, such as nonreciprocal
hopping systems~\cite{Mondal2022,Mondal2023} or models with imaginary
on-site potentials~\cite{Zhou2018,Zhou2021,Das2024,Mondal2024}.
By contrast, DQPTs in non-Hermitian systems under ramp protocols and
without particle-number conservation remain much less explored.

In this work, we study DQPTs in the imbalanced-pairing Kitaev model~\cite{Li2018} driven by a linearly time-dependent chemical potential within the biorthogonal framework. This model is pseudo-Hermitian and possesses time-reversal symmetry rather than combined PT symmetry~\cite{Mostafazadeha,Mostafazadehb,Mostafazadehc,Mostafazadeh2003,Mostafazadeh2010}. Its equilibrium phase diagram consists of a gapped superconducting phase with unbroken time-reversal symmetry, a gapped normal phase, and a gapless non-Hermitian phase with broken time-reversal symmetry.
Our findings reveal that, in non-Hermitian systems, DQPTs under a ramp protocol occur only when the post-ramp Hamiltonian's spectrum is real. 
In addition, we show that, when the parameter that controls the degree of non-Hermiticity $\gamma$ is positive, the dynamics closely resemble those of the Hermitian case.
Moreover, for a ramp crossing two critical or exceptional points, we identify a critical sweep velocity above which DQPTs disappear. This critical velocity decreases as the non-Hermiticity $\gamma$ is decreased and vanishes for staggered pair creation and annihilation, i.e., $\gamma=-1$. Consequently, for $\gamma<-1$, DQPTs are completely suppressed for any sweep velocity.

%-----------------------------
%-----------------------------
\section{Biorthogonal Formulation for Ramp Dynamics}

We consider an integrable system that decomposes into independent two-level Hamiltonians $H_k(h)$ for each momentum mode $k$, where $h$ denotes a tunable control parameter.
The corresponding many-body Hamiltonian is written as
\begin{equation}
\mathcal{H}(h)=\sum_k H_k(h).
\end{equation}
This class of models provides a general framework for investigating equilibrium and dynamical quantum phase transitions and encompasses a broad class of spin-chain and fermionic models.

We assume that, at the initial time $t_i$, the system is prepared in the ground state of the initial Hamiltonian $\mathcal{H}_i=\sum_k H_k(h_i)$,
\begin{equation}
 |\Psi_R^g(h_i)\rangle=\prod_k |\psi_{R,k}^g(h_i)\rangle,
\end{equation}
where $g$ and $e$ denote the ground and excited states, respectively, and the right eigenstates satisfy
\begin{equation}
H_k(h)|\psi_{R,k}^{g,e}(h)\rangle
=\varepsilon_k^{g,e}(h)|\psi_{R,k}^{g,e}(h)\rangle.   
\end{equation}
%
%The corresponding left eigenvectors, $\langle\psi_{L,k}^{g,e}(h)|$, satisfy
%\begin{equation}
%\langle\psi_{L,k}^{g,e}(h)|H_k(h)
%=\varepsilon_k^{g,e}(h)\langle\psi_{L,k}^{g,e}(h)|.
%\end{equation}
%
The corresponding left eigenvectors are defined as left eigenvectors of
the same non-Hermitian Hamiltonian,
\begin{equation}
\langle\psi_{L,k}^{g,e}(h)|H_k(h)
=
\varepsilon_k^{g,e}(h)
\langle\psi_{L,k}^{g,e}(h)| ,
\end{equation}
equivalently, the associated left ket satisfies
\begin{equation}
H_k^\dagger(h)|\psi_{L,k}^{g,e}(h)\rangle
=
\left[\varepsilon_k^{g,e}(h)\right]^*
|\psi_{L,k}^{g,e}(h)\rangle .
\end{equation}
The left and right eigenstates obey the biorthogonality relation
\begin{equation}
\label{biorthogonality-relation}
\bl
\langle\psi_{L,k}^{s}(h)|\psi_{R,k}^{s'}(h)\rangle
=
\delta_{ss'},
\qquad
s,s'\in\{g,e\},
\el
\end{equation}
and the completeness relation
\begin{equation}
\label{completeness-relation}
|\psi_{R,k}^{g}(h)\rangle\langle\psi_{L,k}^{g}(h)|
+
|\psi_{R,k}^{e}(h)\rangle\langle\psi_{L,k}^{e}(h)|
=
1.
\end{equation}

We consider a linear ramp of the control parameter,
\begin{equation}
h(t)=h_f+vt,
\end{equation}
from the initial value $h_i$ at time $t_i$ to the final value $h_f$ at $t_f\rightarrow0^-$. During the ramp, adiabatic evolution breaks down when the system crosses a quantum critical point or an exceptional point at a finite sweep velocity $v$. Consequently, the final state
\begin{equation}
|\Psi_R(h_f)\rangle=\prod_k |\psi_{R,k}(h_f)\rangle
\end{equation}
is, in general, no longer the ground state of the post-ramp many-body Hamiltonian
\begin{equation}
\mathcal{H}_f=\sum_k H_k(h_f).
\end{equation}
Instead, each momentum mode is described by a superposition of the instantaneous ground and excited right eigenstates of $H_k(h_f)$,
\begin{equation}
|\psi_{R,k}(h_f)\rangle
=
v_k(h_f)|\alpha_{R,k}(h_f)\rangle
+
u_k(h_f)|\beta_{R,k}(h_f)\rangle,
\end{equation}
where the expansion coefficients are given by the biorthogonal projections
\begin{equation}
\begin{aligned}
v_k(h_f)
&=
\langle\alpha_{L,k}(h_f)|\psi_{R,k}(h_f)\rangle,
\\
u_k(h_f)
&=
\langle\beta_{L,k}(h_f)|\psi_{R,k}(h_f)\rangle .
\end{aligned}
\end{equation}
Here,
\begin{equation}
|\alpha_{R,k}(h_f)\rangle
\equiv
|\psi^g_{R,k}(h_f)\rangle,
\qquad
|\beta_{R,k}(h_f)\rangle
\equiv
|\psi^e_{R,k}(h_f)\rangle,
\end{equation}
denote the instantaneous ground- and excited-state branches of
$H_k(h_f)$, respectively.

In contrast to Hermitian systems, the right eigenstates of a non-Hermitian Hamiltonian are generally not orthogonal. Therefore, the transition probability cannot be defined simply as
\begin{equation}
p_k=\left|\langle\beta_{R,k}(h_f)|\psi_{R,k}(h_f)\rangle\right|^2 .
\end{equation}
To construct the biorthogonal Loschmidt amplitude, the final right state must
be paired with its dual state in the left-eigenvector basis. We therefore
introduce the corresponding biorthogonal partner
\begin{equation}
|\tilde{\Psi}(h_f)\rangle
=
\prod_k|\tilde{\psi}_k(h_f)\rangle,
\end{equation}
which is not the Hermitian conjugate of $|\Psi_R(h_f)\rangle$.
Following Ref.~\cite{Deng2025}, the
single-mode dual state is written as
\begin{equation}
\begin{aligned}
&
|\tilde{\psi}_{k}(h_f)\rangle
=
\\
&
v_k(h_f)\frac{N^g_{R,k}(h_f)}{N^g_{L,k}(h_f)}
|\alpha_{L,k}(h_f)\rangle 
\!+\!
u_k(h_f)\frac{N^e_{R,k}(h_f)}{N^e_{L,k}(h_f)}
|\beta_{L,k}(h_f)\rangle .
\end{aligned}
\end{equation}
Here
\begin{equation}
|\alpha_{L,k}(h_f)\rangle
\equiv
|\psi^g_{L,k}(h_f)\rangle,
\qquad
|\beta_{L,k}(h_f)\rangle
\equiv
|\psi^e_{L,k}(h_f)\rangle ,
\end{equation}
are the left eigenstates associated with the ground and excited branches.
The normalization factors are defined as
\begin{equation}
\begin{aligned}
N^g_{R,k} &= \sqrt{\langle\alpha_{R,k}|\alpha_{R,k}\rangle},
\qquad
N^e_{R,k} = \sqrt{\langle\beta_{R,k}|\beta_{R,k}\rangle},\\
N^g_{L,k} &= \sqrt{\langle\alpha_{L,k}|\alpha_{L,k}\rangle},
\qquad
N^e_{L,k} = \sqrt{\langle\beta_{L,k}|\beta_{L,k}\rangle},
\end{aligned}
\end{equation}
where all states are evaluated at $h_f$.
\medskip

Following the biorthogonal construction of
Ref.~\cite{Jing2024}, we factorize the normalized mode-resolved
Loschmidt echo into two reciprocal biorthogonal amplitudes,
\begin{equation}
\label{eq:forward-amplitude}
{\cal G}_{k}(t)
=
\frac{
\langle\tilde{\psi}_{k}(h_f)|\psi_{R,k}(t)\rangle
}{
\sqrt{
\langle\tilde{\psi}_{k}(h_f)|\psi_{R,k}(h_f)\rangle
\langle\tilde{\psi}_{k}(t)|\psi_{R,k}(t)\rangle
}
},
\end{equation}
and
\begin{equation}
\label{eq:backward-amplitude}
\widetilde{\cal G}_{k}(t)
=
\frac{
\langle\tilde{\psi}_{k}(t)|\psi_{R,k}(h_f)\rangle
}{
\sqrt{
\langle\tilde{\psi}_{k}(h_f)|\psi_{R,k}(h_f)\rangle
\langle\tilde{\psi}_{k}(t)|\psi_{R,k}(t)\rangle
}
}.
\end{equation}
These two amplitudes describe the forward and reciprocal biorthogonal
overlaps between the post-ramp state and its time-evolved counterpart.
Their product defines the normalized mode-resolved biorthogonal
Loschmidt echo~\cite{Jing2024,Wang2026b,Gu2026,Fu2025},
\begin{equation}
\label{eq:Loschmidt-echo}
{\cal L}_{k}(t)
=
{\cal G}_{k}(t)\widetilde{\cal G}_{k}(t)
=
\frac{
\langle\tilde{\psi}_{k}(h_f)|\psi_{R,k}(t)\rangle
\langle\tilde{\psi}_{k}(t)|\psi_{R,k}(h_f)\rangle
}{
\langle\tilde{\psi}_{k}(h_f)|\psi_{R,k}(h_f)\rangle
\langle\tilde{\psi}_{k}(t)|\psi_{R,k}(t)\rangle
}.
\end{equation}
The many-body biorthogonal Loschmidt echo is
\begin{equation}
{\cal L}(t)=\prod_k{\cal L}_k(t),
\end{equation}
and the associated return rate is given by
\begin{equation}
g(t)
=
-\frac{1}{N}\ln{\cal L}(t)
=
-\frac{1}{N}\sum_k\ln{\cal L}_k(t),
\end{equation}
where $N$ denotes the system size.
The time-evolved right state and its biorthogonal partner are given by
\begin{equation}
\bl
|\psi_{R,k}(t)\rangle
&=
e^{-iH_k(h_f)t}|\psi_{R,k}(h_f)\rangle,
\\
|\tilde{\psi}_{k}(t)\rangle
&=
e^{-iH_k^{\dagger}(h_f)t}|\tilde{\psi}_{k}(h_f)\rangle.
\el
\end{equation}
Using the expansion of the final state in the biorthogonal eigenbasis, the
two mode-resolved Loschmidt amplitudes become
\begin{equation}
\begin{aligned}
{\cal G}_k(t)
&=
[(1-p_k)+p_k e^{-i\delta\varepsilon_k t}
]
e^{-i\varepsilon_k^{g}(h_f)t},
\\
\widetilde{\cal G}_k(t)
&=
[(1-p_k)+p_k e^{i\delta\varepsilon_k t}
]
e^{i\varepsilon_k^{g}(h_f)t},
\end{aligned}
\end{equation}
where
$\delta\varepsilon_k
=
\varepsilon_k^e(h_f)-\varepsilon_k^g(h_f)$.
Their product therefore gives
\begin{equation}
{\cal L}_k(t)
=
1-4p_k(1-p_k)
\sin^2\left(\frac{\delta\varepsilon_k t}{2}\right).
\end{equation}
The corresponding normalized transition probability to the excited state is given by~\cite{Deng2025}
\begin{equation}
\bl
p_k
&=
\frac{
\langle\tilde{\psi}_{k}(h_f)|\beta_{R,k}(h_f)\rangle
\langle\beta_{L,k}(h_f)|\psi_{R,k}(h_f)\rangle}
{
\langle\tilde{\psi}_{k}(h_f)|\psi_{R,k}(h_f)\rangle
\langle\beta_{L,k}(h_f)|\beta_{R,k}(h_f)\rangle}
\\
\label{eq4}
&=
\frac{|u_k(h_f)|^2}
{
\Big[
\frac{
N^g_{R,k}(h_f)N^e_{L,k}(h_f)}
{
N^g_{L,k}(h_f)N^e_{R,k}(h_f)}
\Big]
|v_k(h_f)|^2
+
|u_k(h_f)|^2
},
\el
\end{equation}
where $u_k(h_f)$ and $v_k(h_f)$ are the expansion coefficients defined above.
It is worth emphasizing that the transition probability defined above
within the biorthogonal framework corresponds to the transition probability to the excited state of the left eigenvectors of Hamiltonian.
In the thermodynamic limit, the dynamical free-energy density is given by
\begin{equation}
\label{eq5}
g(t)
=
-\frac{1}{2\pi}
\int_0^\pi
\ln\left[
1-4p_k(1-p_k)
\sin^2\left(\frac{\delta\varepsilon_k t}{2}\right)
\right] dk .
\end{equation}

The dynamical free-energy density becomes nonanalytic when the argument of the logarithm in Eq.~(\ref{eq5}) vanishes, yielding the critical times
%
%%%%%%%%%%%%%%%%%%%%%%%%%%%%  Eq. 6 %%%%%%%%%%%%%%%%%%%%%%%%%%%%%%%%
\bea
\label{eq6}
t_n^*=\frac{\pi}{\delta\varepsilon_{k^*}}(2n+1),
\qquad
n=0,1,2,\cdots.
\eea
%%%%%%%%%%%%%%%%%%%%%%%%%%%%%%%%%%%%%%%%%%%%%%%%%%%%%%%%%%%%%%%%%%%
%
These critical times exist only when the post-ramp Hamiltonian possesses a real energy spectrum, i.e.,
$\varepsilon_k^{g,e}(h_f)\in\mathbb{R}$.
They occur at the critical momentum $k^*$ satisfying $p_{k^*}=1/2$.
Therefore, similar to the sudden-quench case in non-Hermitian
systems~\cite{Wang2026b}, DQPTs emerge when the ramp ends in the
time-reversal-symmetric phase.
For the imbalanced-pairing Kitaev model, the explicit Landau-Zener solution
leading to the excitation probability used throughout this work is presented
in Sec.~\ref{Analytical-Solution-Ramp-Dynamics}. The symmetry properties and the relevant exceptional-point
topology are summarized in Appendices~\ref{app:symmetry}
and~\ref{app:nh-topology}, respectively.

In addition, the dynamical topological order parameter (DTOP) provides a topological characterization of DQPTs~\cite{Budich2016,Bhattacharya,Yang2018,Tang2024,Wang2024,Guo2020}.
The DTOP is quantized, and changes by one unit at the critical times, signaling the topological nature of the transition.
It is constructed from the gauge-invariant Pancharatnam geometric phase associated with the Loschmidt amplitude~\cite{Budich2016,Bhattacharya} and, for non-Hermitian systems, is defined as~\cite{Zhou2018,Jing2024,Wang2026b,Gu2026,Fu2025}
\begin{eqnarray}
\label{eq7}
N_w=\frac{1}{2\pi}\int_0^\pi\frac{\partial\phi^G(k,t)}{\partial k}\mathrm{d}k.
\end{eqnarray}
The geometric phase is obtained by subtracting the dynamical phase from the total phase,
\begin{equation}
\phi^G(k,t)=\phi(k,t)-\phi^D(k,t).
\end{equation}
The total phase is obtained from the polar decomposition of the
mode-resolved biorthogonal Loschmidt amplitude,
\begin{equation}
{\cal G}_k(t)
=
|{\cal G}_k(t)|e^{i\phi(k,t)},
\end{equation}
while the dynamical phase is given by
\begin{equation}
\no
\bl
\phi^{D}(k,t) =
& -\int_{0}^{t}dt'
\frac{\langle\tilde{\psi}_{k}(t')|H_k(h_f)|\psi_{R,k}(t')\rangle}
{\langle\tilde{\psi}_{k}(t')|\psi_{R,k}(t')\rangle}
\\
&+\frac{i}{2}\ln\left[
\frac{\langle\tilde{\psi}_{k}(t)|\psi_{R,k}(t)\rangle}
{\langle\tilde{\psi}_{k}(t_f)|\psi_{R,k}(t_f)\rangle}
\right].
\el
\end{equation}
For the present two-level system, the total and dynamical phases are given by
\bea
\no
\phi(k,t)=
\tan^{-1}
\left[
\frac{-p_k\sin(\delta\varepsilon_k t)}
{(1-p_k)+p_k\cos(\delta\varepsilon_k t)}
\right]
-\varepsilon_k^{g}(h_f)t,
\eea
and
\bea
\no
\phi^{D}(k,t)=-p_k\delta\varepsilon_k t
-\varepsilon_k^{g}(h_f)t,
\eea
so that
\begin{equation}
    \no
    \bl
\phi_k^G =
\tan^{-1}
\left[
\frac{-p_k\sin(\delta\varepsilon_k t)}
{(1-p_k)+p_k\cos(\delta\varepsilon_k t)}
\right]
+p_k\delta\varepsilon_k t.
\el
\end{equation}
For comparison, the conventional return rate is defined as
\begin{equation}
g^{co}(t)=
-\frac{1}{2\pi}\int_0^\pi
\ln\left|
\langle\psi_{R,k}(h_f)|\psi_{R,k}(t)\rangle
\right|^2 dk,
\end{equation}
where $\langle\psi_{R,k}(h_f)|$ denotes the Hermitian conjugate of
$|\psi_{R,k}(h_f)\rangle$.

The above formulation is completely general and applies to any integrable two-band pseudo-Hermitian Hamiltonian. In the following, we specialize it to the imbalanced-pairing Kitaev model.

%%%%%%%%%%%%%%%%%%%%%%%%%%%%%%%%%%%%%%%%%%%%%%%%%%%%%%%%%%%%%%%%%%%%%%%%%%%%%%
\section{Model and Phase Diagram}
%%%%%%%%%%%%%%%%%%%%%%%%%%%%%%%%%%%%%%%%%%%%%%%%%%%%%%%%%%%%%%%%%%%%%%%%%%%%%%

The imbalanced pairing Kitaev model, which serves as a prototypical
pseudo-Hermitian superconducting system~\cite{Li2018}, is described by
\begin{equation}
\label{Hamiltonian}
\bl
\mathcal{H}(t)
=
&
\frac{w}{2}\sum_{j=1}^{N}
\left(c_{j}^{\dagger}c_{j+1}+{\rm H.c.}\right)
-\mu(t)\sum_{j=1}^{N}c_{j}^{\dagger}c_j
\\
&
-
\frac{1}{2}
\sum_{j=1}^{N}
\left(
\Delta c_{j}^{\dagger}c_{j+1}^{\dagger}
+
\gamma\Delta^* c_{j+1}c_j
\right),
\el
\end{equation}
where $c_j^\dagger$ ($c_j$) creates (annihilates) a fermion at site $j$,
$w$ is the nearest-neighbor hopping amplitude, $\mu(t)$ is the
time-dependent chemical potential, $\Delta$ is the $p$-wave pairing
amplitude, and $\gamma$ controls the degree of non-Hermiticity. 
Without loss of generality, the global superconducting phase is fixed such
that $\Delta$ is real. We also take the imbalance parameter $\gamma$ to be
real throughout.
We impose anti-periodic boundary conditions,
$c_{N+1}=-c_1$, which lead to the allowed positive momenta
$
k=(2m-1)\pi/N,
\;
m=1,2,\ldots,{N}/{2}.
$
The momenta $k=0$ and $k=\pi$ are excluded for finite $N$ and are
recovered only as limiting values in the thermodynamic limit.
The Hamiltonian is invariant under the antiunitary time-reversal operator
$\mathcal{T}$ satisfying
$\mathcal{T}i\mathcal{T}^{-1}=-i$~\cite{Li2018}. 
The Hermitian limit is recovered at $\gamma=1$, where the Hamiltonian
additionally possesses parity-time ($\mathcal{PT}$) symmetry.
The case $\gamma=-1$ corresponds to staggered pair creation and annihilation and possesses distinct spectral properties. 
The symmetry properties and their connection to pseudo-Hermiticity are summarized in
Appendix~\ref{app:symmetry}.

Using the Fourier transformation 
\begin{equation}
c_j=\frac{1}{\sqrt{N}}e^{-i\pi/4}\sum_k e^{ikj}c_k,
\quad
c_j^\dagger=\frac{1}{\sqrt{N}}e^{i\pi/4}\sum_k e^{-ikj}c_k^\dagger,
\end{equation}
the Hamiltonian decomposes into independent momentum sectors. Up to an additive constant and after grouping the $\pm k$ sectors, we can write the Hamiltonian as:
\begin{equation}
\mathcal{H}(t)=\sum_{k>0}\Psi_k^\dagger H_k(t)\Psi_k,
\end{equation}
where the Bogoliubov-de Gennes Hamiltonian in the Nambu basis $\Psi_k=(c_k,c_{-k}^\dagger)^T$ is given by
\begin{equation}
\label{eq11}
H_k(t)=\begin{pmatrix}
w\cos k-\mu(t) & \Delta\sin k \\ \gamma\Delta\sin k & \mu(t)-w\cos k
\end{pmatrix}.
\end{equation}
Equivalently,
\begin{equation}
H_k(t) = d_x(k)\sigma_x + i d_y(k)\sigma_y + d_z(k)\sigma_z,
\end{equation}
with
\begin{equation}
d_x= \frac{1+\gamma}{2}\Delta\sin k, \quad d_y=\frac{1-\gamma}{2}\Delta\sin k, \quad d_z=w\cos k-\mu(t).
\end{equation}
For a time-independent chemical potential, $\mu(t)=\mu$, the quasiparticle
spectrum follows from $\det[H_k-\varepsilon_k I]=0$, which yields the quasiparticle energies as
\begin{equation}
\label{quasiparticle-spectrum}
\varepsilon_k^{g}=-\varepsilon_k^{e}=-\varepsilon_k=-\sqrt{(w\cos k-\mu)^2+\gamma\Delta^2\sin^2k}. 
\end{equation}
The spectrum is real in the time-reversal-symmetric regime and becomes
complex after crossing the exceptional points discussed in
Appendix~\ref{app:symmetry}. In the complex-spectrum regime, the labels $g$ and $e$ denote analytic continuations of the negative- and positive-energy branches, rather than ground and excited states in the strict energetic sense.

Introducing the compact notation 
\begin{equation}
\mu_k(t)=w\cos k-\mu(t), 
\qquad \Delta_k=\Delta\sin k, 
\end{equation} 
together with 
\begin{equation} 
a_k(t)=\mu_k(t)-\varepsilon_k(t), \end{equation} 
the biorthogonal normalization product is given by
\begin{equation}
\mathcal{B}_k(t)=a_k^2(t)+\gamma\Delta_k^2.
\end{equation}
%
%In the broken-symmetry phase, $\mathcal{B}_k(t)$ is generally complex and, as a biorthogonal inner product, should not be interpreted as a positive-definite Hermitian norm.
%
%All following eigenvector formulas are valid away from exceptional points, where $\mathcal{B}_k(t) \neq 0$ and the Hamiltonian is diagonalizable. 
%
%Assuming real $\gamma$ and $\Delta_k$, the properly normalized instantaneous right kets and left bras are constructed directly as:
In the broken-symmetry phase, $\mathcal{B}_k(t)$ is generally complex
and, as a biorthogonal normalization product, should not be interpreted
as a positive-definite Hermitian norm. In the parameter region where
$\mathcal{B}_k(t)\neq0$, the instantaneous right eigenvectors and left
eigenvectors can be chosen as
\begin{equation}
\bl
|\psi_{R,k}^{g}(t)\rangle &=  \frac{1}{\sqrt{\mathcal{B}_k(t)}}
\begin{pmatrix} 
a_{k}(t) \\  \gamma\Delta_k 
\end{pmatrix}, 
\\  |\psi_{R,k}^{e}(t)\rangle &=  \frac{1}{\sqrt{\mathcal{B}_k(t)}}
\begin{pmatrix} 
-\Delta_k \\  a_{k}(t) 
\end{pmatrix}, 
\\ \langle\psi_{L,k}^{g}(t)| &=  \frac{1}{\sqrt{\mathcal{B}_k(t)}}
\begin{pmatrix} 
a_{k}(t) & \Delta_k 
\end{pmatrix}, 
\\  \langle\psi_{L,k}^{e}(t)| &=  \frac{1}{\sqrt{\mathcal{B}_k(t)}}
\begin{pmatrix} 
-\gamma\Delta_k & a_{k}(t) 
\end{pmatrix}.
\el
\end{equation}
This choice of eigenvectors becomes singular when
$\mathcal{B}_k(t)=0$. Such a singularity may occur either at an
exceptional point or at isolated momenta where this eigenvector
parametrization vanishes. In the latter case, the Hamiltonian remains
diagonalizable, and an alternative nonsingular eigenvector gauge must
be adopted.
The associated left kets, used in the construction of the dual state in the main text, are the Hermitian conjugates of the above left bras:
\begin{equation}
\bl
|\psi^g_{L,k}(t)\rangle &= \frac{1}{\sqrt{\mathcal{B}_k^*(t)}}
\begin{pmatrix} 
a^*_{k}(t) \\ \Delta_k 
\end{pmatrix}, 
\\ 
|\psi^e_{L,k}(t)\rangle &= \frac{1}{\sqrt{\mathcal{B}_k^*(t)}}
\begin{pmatrix} 
-\gamma\Delta_k \\ a^*_{k}(t) 
\end{pmatrix}.
\el
\end{equation}
With these definitions, the left bras and right kets strictly satisfy the
biorthogonality relation, Eq.~(\ref{biorthogonality-relation}),
%\begin{equation}
%\langle\psi_{L,k}^{s}(t) | \psi_{R,k}^{s'}(t) \rangle = \delta_{ss'}, \qquad s,s'\in\{g,e\},
%\end{equation}
together with the completeness relation, Eq.~(\ref{completeness-relation}).
%\begin{equation}
%|\psi_{R,k}^{g}(t)\rangle \langle\psi_{L,k}^{g}(t)| + |\psi_{R,k}^{e}(t)\rangle \langle\psi_{L,k}^{e}(t)| = 1.
%\end{equation}
Finally, to evaluate the biorthogonal Loschmidt echo and the transition probability $p_k$, one requires the standard Hermitian norms of these basis vectors. Because $a_{k}(t)$  can become complex in the broken-symmetry phase, one must strictly use the absolute square to ensure real-valued, positive-definite norms. Using our properly normalized biorthogonal basis, the Hermitian normalization factors evaluate analytically to:
\begin{equation}
\begin{aligned}
N^{g}_{R,k}(t)
=
N^{e}_{L,k}(t)
&=
\left[
\frac{|a_k(t)|^2+|\gamma\Delta_k|^2}
{|\mathcal{B}_k(t)|}
\right]^{1/2},
\\
N^{e}_{R,k}(t)
=
N^{g}_{L,k}(t)
&=
\left[
\frac{|a_k(t)|^2+|\Delta_k|^2}
{|\mathcal{B}_k(t)|}
\right]^{1/2}.
\end{aligned}
\end{equation}

%
%%%%%%%%%%%%%%%%%%%%%%%  Fig.1   %%%%%%%%%%%%%%%%%%%%%%%
\begin{figure}[t]
\centerline{\includegraphics[width=1\linewidth]{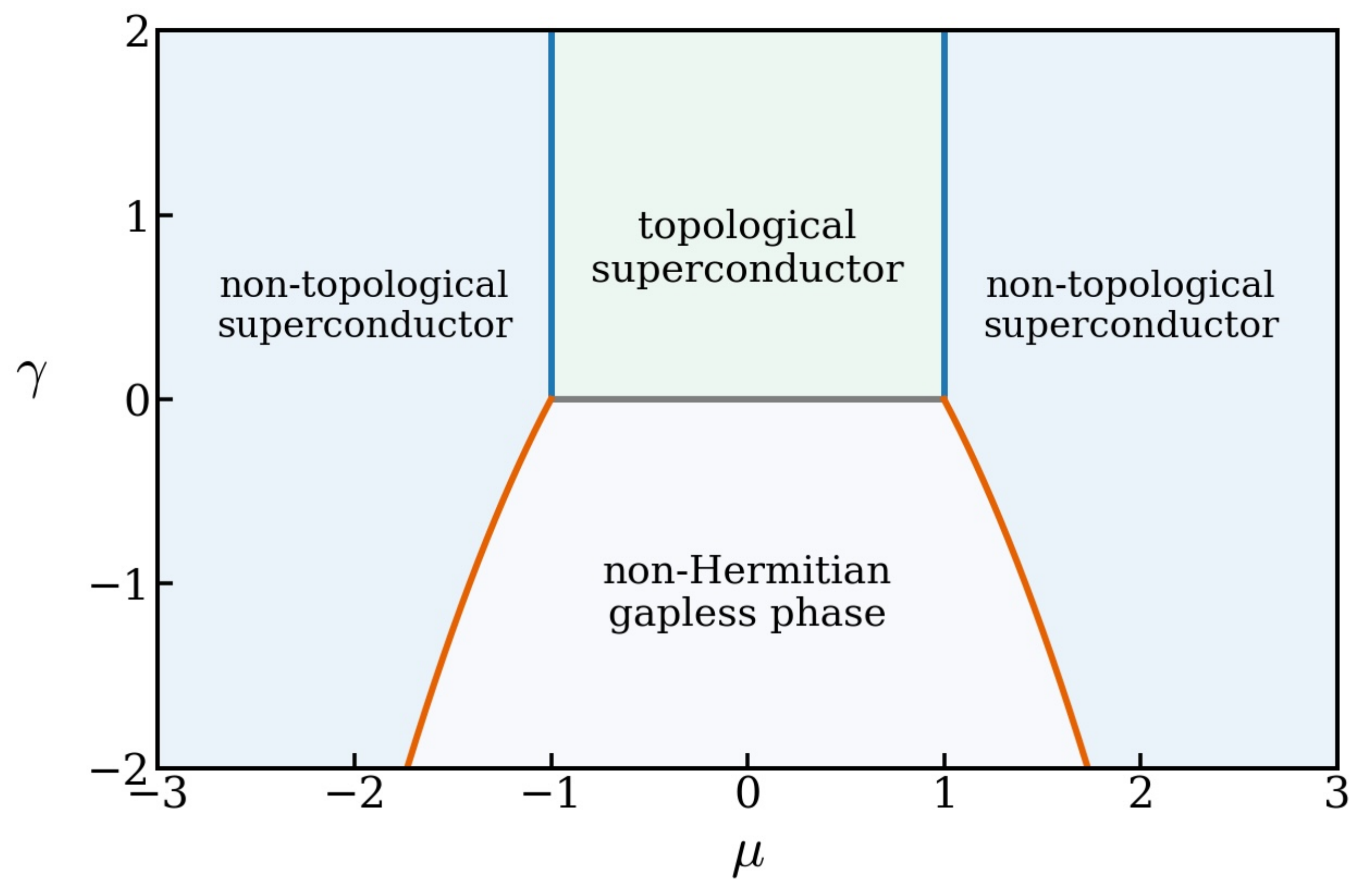}}
\caption{
Equilibrium phase diagram of the imbalanced-pairing Kitaev model in the
$\mu$-$\gamma$ plane. The diagram shows the topological superconducting,
non-topological superconducting, and gapless non-Hermitian phases. 
The boundaries of the gapless non-Hermitian region for $\gamma<0$ correspond to
exceptional points.}
\label{fig1}
\end{figure}
%%%%%%%%%%%%%%%%%%%%%%%%%%%%%%%%%%%%%%%%%%%%%%%%%%%%%%%
%

Before turning to the nonequilibrium dynamics, we briefly summarize the
equilibrium phase diagram. Unless stated otherwise, we set
$w=\Delta=1$. The phase boundaries follow directly from the quasiparticle
spectrum in Eq.~(\ref{quasiparticle-spectrum}). For $\gamma>0$, the spectrum
is real for all momenta, and the gap closes at $\mu=\pm w$, marking the
transition between the topological and non-topological superconducting
phases. For $\gamma<0$, by contrast, the spectrum becomes complex within a
finite region bounded by the exceptional-point condition
\begin{equation}
\label{exceptional-point-condition}
(w\cos k-\mu)^2+\gamma\Delta^2\sin^2k=0.
\end{equation}
For $w=\Delta=1$, this condition gives the phase boundaries
\begin{equation}
\mu=\pm\sqrt{1-\gamma},
\end{equation}
which separate the gapped superconducting phases from the gapless
non-Hermitian phase shown in Fig.~\ref{fig1}.

We now investigate the nonequilibrium dynamics generated by a linear ramp of
the chemical potential. Since the momentum sectors evolve independently, the
dynamics can be solved exactly using the Landau-Zener formalism developed in
Sec.~\ref{Analytical-Solution-Ramp-Dynamics}. The symmetry properties of the model and their relation to
pseudo-Hermiticity are discussed in Appendix~\ref{app:symmetry}.

\section{Analytical Solution of the Ramp Dynamics}
\label{Analytical-Solution-Ramp-Dynamics}
For a linear ramp of the chemical potential,
$\mu(t)=\mu_f+vt$, each momentum sector evolves independently.
Using Eq.~(\ref{eq11}), the Hamiltonian can be written in the
Landau-Zener form
\begin{equation}
\label{eq:LZ-Hamiltonian}
H_k(t)
=
\left(
\begin{array}{cc}
-v\tau_k(t)  & \Delta_k \\
\gamma \Delta_k & v\tau_k(t)
\end{array}
\right),
\end{equation}
where $\tau_k(t)=-\mu_k(t)/v$.
%
%\begin{equation}
%\label{eq:tau-k}
%\tau_k(t)
%=
%\frac{1}{v}[\mu(t)-w\cos k],
%\qquad
%\Delta_k=\Delta\sin k .
%\end{equation}
%
The excitation probabilities can therefore be obtained exactly by solving
the time-dependent Schr\"odinger equation~\cite{Torosov2017}
\begin{equation}
\label{eq:tdse}
i\frac{d}{dt}|\psi_{R,k}(t)\rangle
=
H_k(t)|\psi_{R,k}(t)\rangle ,
\end{equation}
with 
the initial condition
\begin{equation}
|\psi_{R,k}(\mu_i)\rangle
=
|\alpha_{R,k}(\mu_i)\rangle .
\end{equation}
Considering 
\begin{equation}
\label{finalState}
|\psi_{R,k}(t)\rangle
=
\begin{pmatrix} 
C_{1}(t) \\ C_{2}(t) 
\end{pmatrix}
=
\begin{pmatrix}
U_{11}(t) & U_{12}(t)\\
U_{21}(t) & U_{22}(t)
\end{pmatrix}
|\alpha_{R,k}(\mu_i)\rangle, 
\end{equation}
the time-evolved state is obtained as

\begin{equation}
\begin{aligned}
C_1(t)&=\frac{U_{11}(t)a_{k}(\mu_i)+U_{12}(t)\gamma\Delta_k}{\sqrt{\mathcal{B}_k(\mu_i)}},
\\
C_2(t)&=\frac{U_{21}(t)a_{k}(\mu_i)+U_{22}(t)\gamma\Delta_k}{\sqrt{\mathcal{B}_k(\mu_i)}} .
\end{aligned}
\end{equation}

%Here
%\begin{equation}
%a_{R,k}(\mu_i)=w\cos k-\mu_i-\varepsilon_k(\mu_i),
%\end{equation}
%and $\mathcal{B}_k(\mu_i) = a_{R,k}^2(\mu_i) + \gamma\Delta_k^2$ is the biorthogonal normalization product of the initial state, as defined in Appendix~\ref{app:diagonalization}.

The time-evolution matrix elements $U_{ij}(t)$ are given by~\cite{Torosov2017}
\begin{equation}
\label{eq:U-elements}
\begin{aligned}
U_{11}(t)
&=
\frac{\Gamma(1-\omega)}{\sqrt{2\pi}}
\Big[
D_{\omega-1}(-z_i)D_{\omega}(z_f)
+
D_{\omega-1}(z_i)D_{\omega}(-z_f)
\Big],
\\
U_{12}(t)
&=
\frac{\Gamma(1-\omega)}
{\lambda\sqrt{\gamma\pi}}
e^{i\pi/4}
\Big[
D_{\omega}(z_i)D_{\omega}(-z_f)
-
D_{\omega}(-z_i)D_{\omega}(z_f)
\Big],
\\
U_{21}(t)
&=
\frac{\lambda\sqrt{\gamma}\,\Gamma(1-\omega)}
{2\sqrt{\pi}}
e^{-i\pi/4}
\Big[
D_{\omega-1}(z_i)D_{\omega-1}(-z_f)
\\&\hspace{4cm}
-
D_{\omega-1}(-z_i)D_{\omega-1}(z_f)
\Big],
\\
U_{22}(t)
&=
\frac{\Gamma(1-\omega)}{\sqrt{2\pi}}
\Big[
D_{\omega}(-z_i)D_{\omega-1}(z_f)
+
D_{\omega}(z_i)D_{\omega-1}(-z_f)
\Big].
\end{aligned}
\end{equation}
The auxiliary quantities appearing in Eq.~(\ref{eq:U-elements}) are
\begin{equation}
\no
\omega=\frac{i\lambda^2}{2},
\qquad
\lambda=\Delta_k\sqrt{\frac{\gamma}{v}},
\qquad
z_{i,f}
=
\sqrt{2v}\,e^{-i\pi/4}\tau_k(t_{i,f}) .
\end{equation}
Here, $\Gamma(\nu)$ denotes the Euler gamma function, and
$D_{\nu}(z)$ denotes the parabolic cylinder function of order $\nu$~\cite{szego1954,abramowitz1988}. 
\medskip

%
%%%%%%%%%%%%%%%%%%%%%%%  Fig.2   %%%%%%%%%%%%%%%%%%%%%%%
\begin{figure*}
\begin{minipage}{\linewidth}
\centerline{\includegraphics[width=0.33\linewidth]{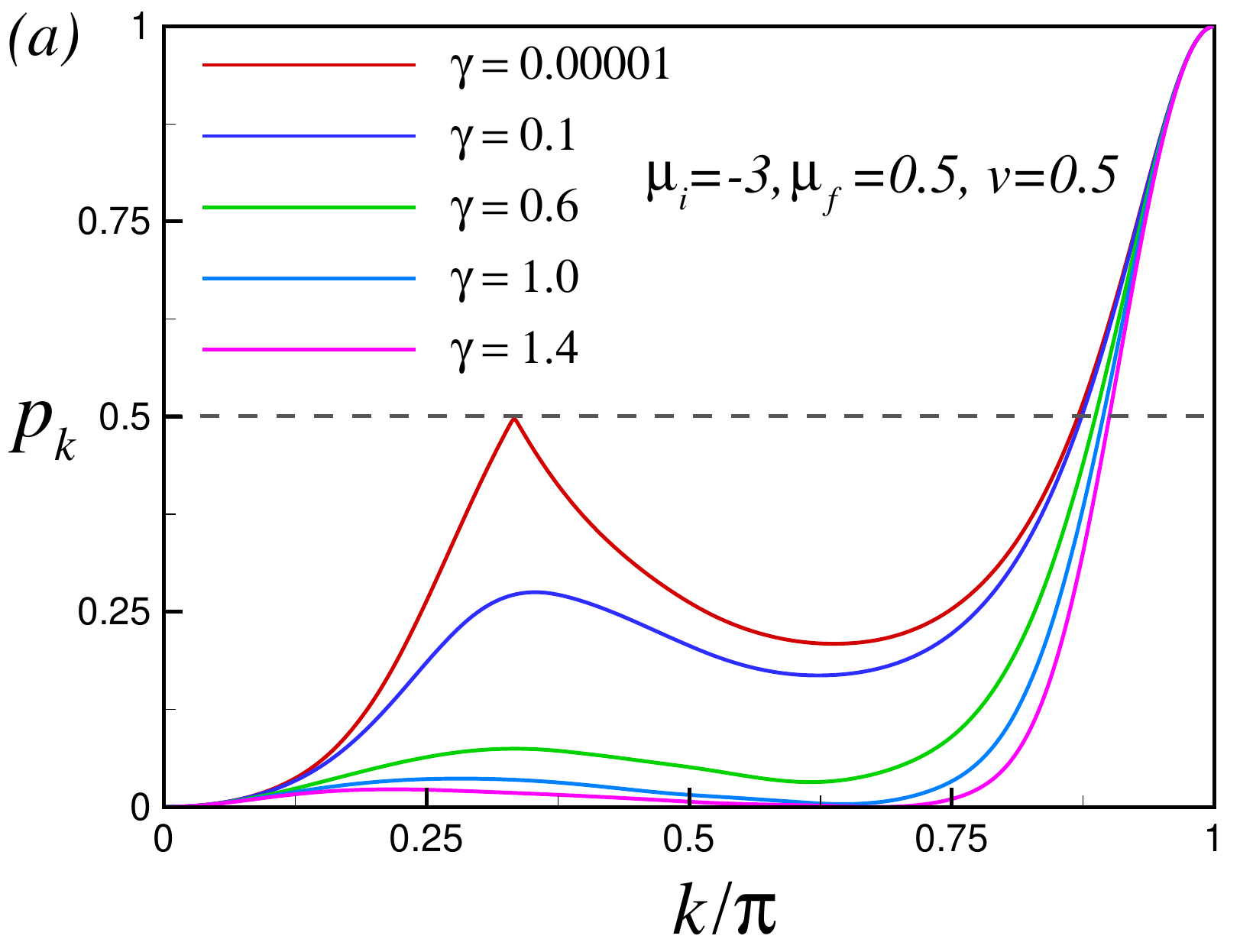}
\includegraphics[width=0.33\linewidth]{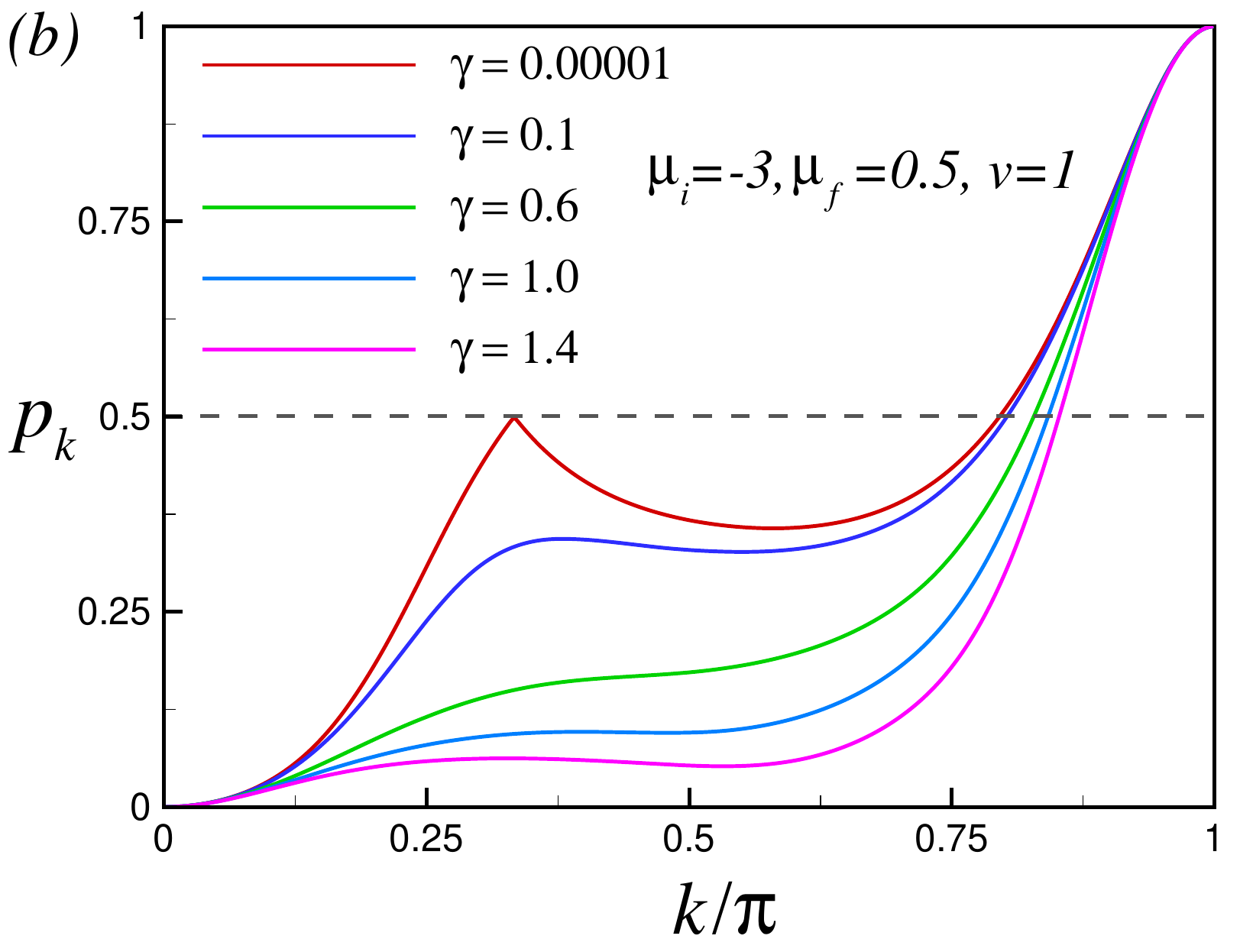}
\includegraphics[width=0.33\linewidth]{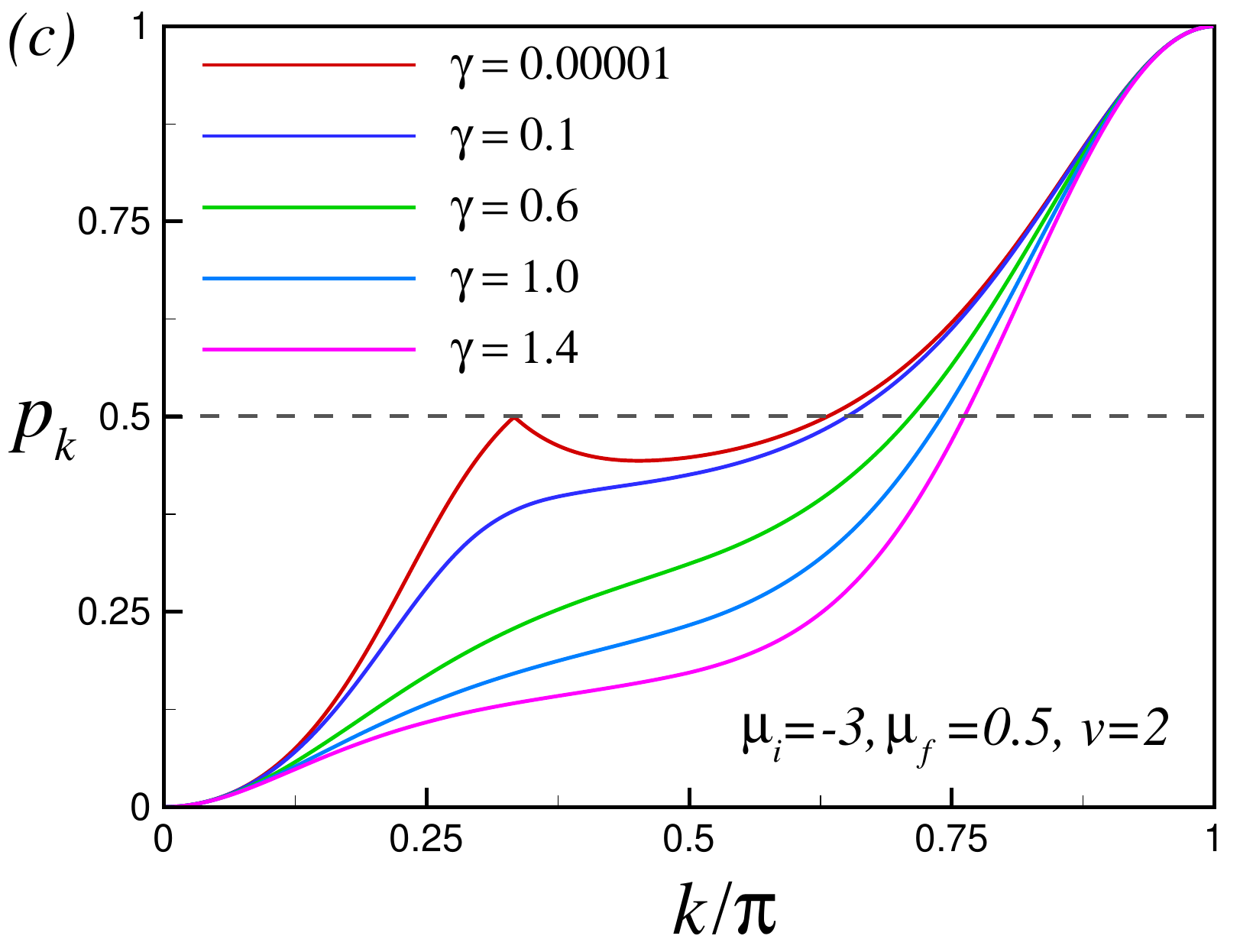}}
\centering
\end{minipage}
%===============================================================
\begin{minipage}{\linewidth}
\centerline{\includegraphics[width=0.31\linewidth]{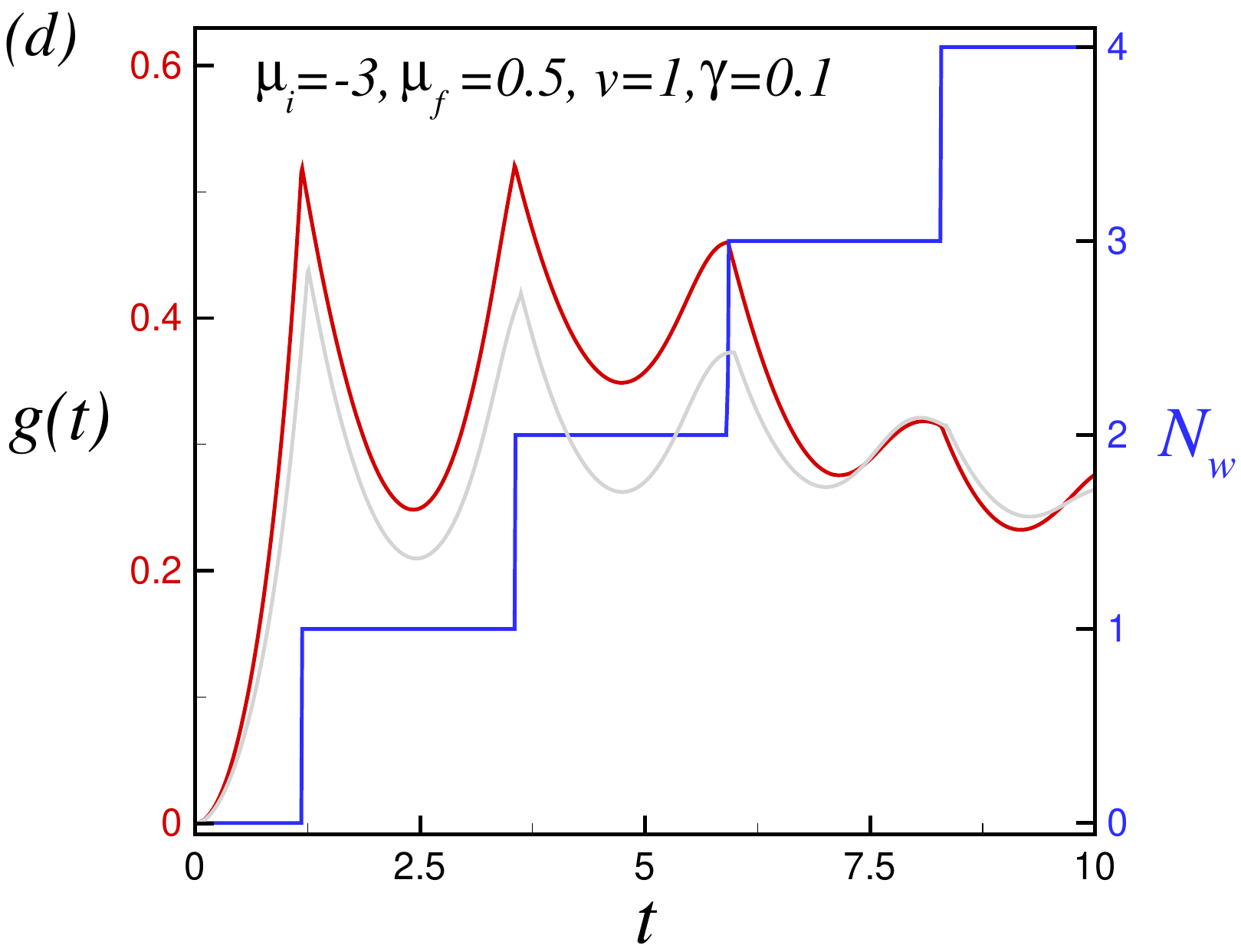}
\hfill
\includegraphics[width=0.31\linewidth]{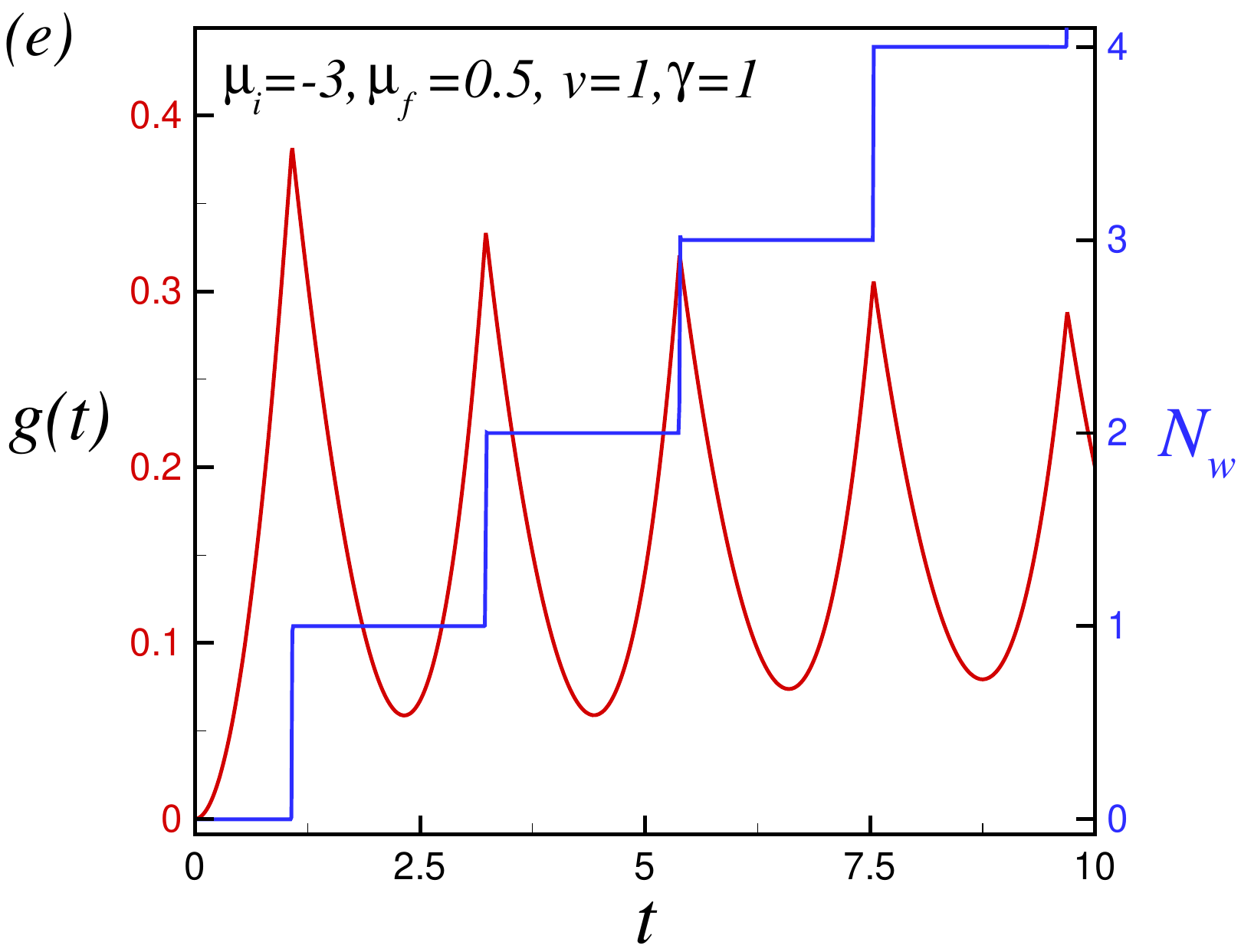}
\hfill
\includegraphics[width=0.31\linewidth]{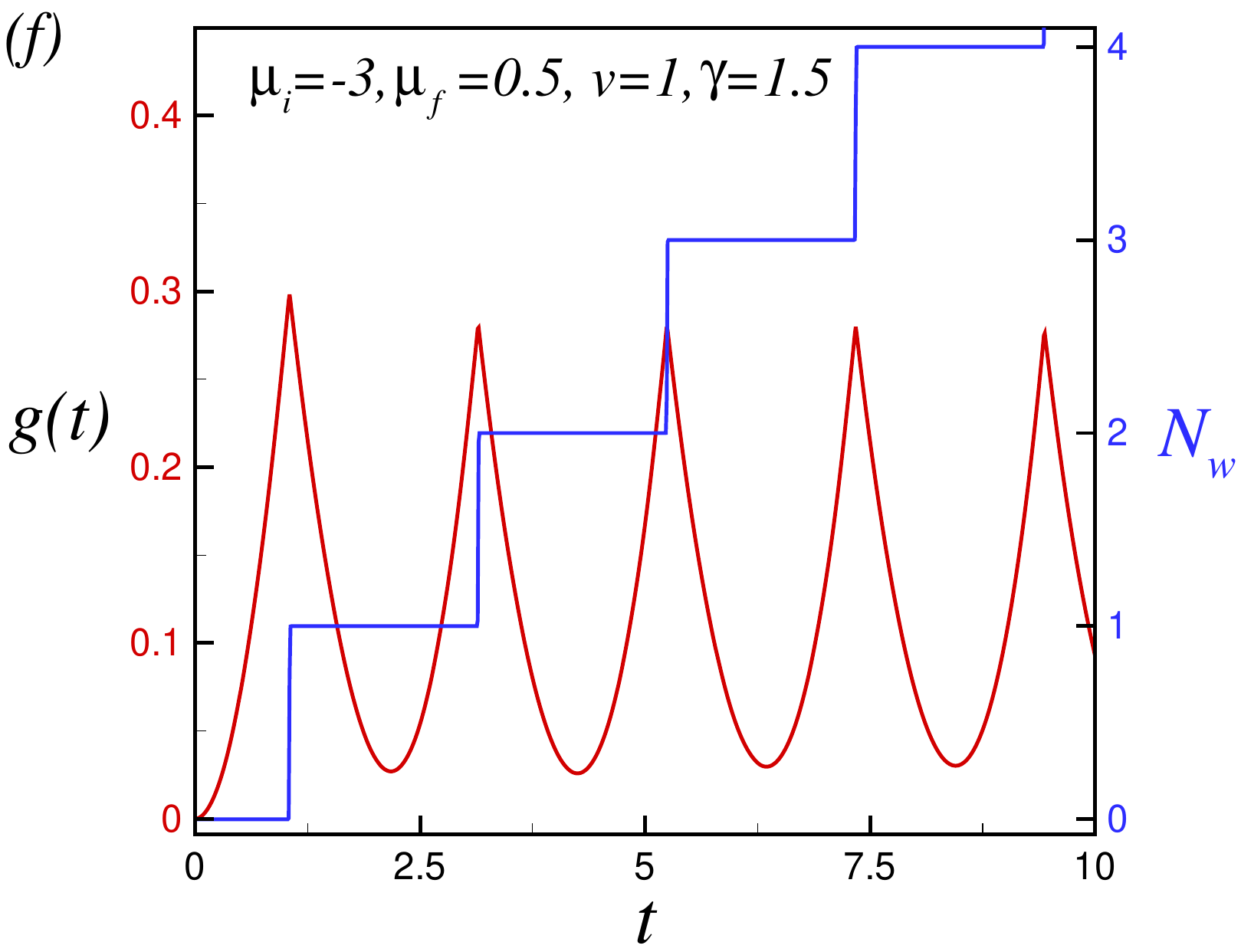}}
\centering
\end{minipage}
%===============================================================
\caption{Transition probability $p_k$ as a function of momentum for ramps from
$\mu_i=-3$ to $\mu_f=0.5$, crossing the single critical point
$\mu_c=-1$. Panels (a)--(c) show $p_k$ for different values of $\gamma$ and
sweep velocities: (a) $v=0.5$, (b) $v=1$, and (c) $v=2$. Panels (d)--(f) show
the corresponding biorthogonal dynamical free energy $g(t)$ and DTOP $N_w(t)$
for $v=1$: (d) $\gamma=0.1$, (e) $\gamma=1$, and (f) $\gamma=1.5$. The blurred
curve in panel (d) denotes the conventional dynamical free energy, shown for
comparison. The system size is $N=1000$.}
\label{fig2}
\end{figure*}
%%%%%%%%%%%%%%%%%%%%%%%%%%%%%%%%%%%%%%%%%%%%%%%%%%%%%%%
%

The state at the end of the ramp,
$|\psi_{R,k}(\mu_f)\rangle$, is obtained from
Eq.~(\ref{finalState}), and the corresponding normalized excitation
probability is then evaluated using Eq.~(\ref{eq4}).
In the asymptotic Landau-Zener limit
$\mu_i\to-\infty$ and $\mu_f\to+\infty$,
where 
$$
|\phi_{R,\bm{k}}(\mu_i)\rangle
=
\begin{pmatrix} 
0 \\ 1 
\end{pmatrix}, 
\quad
|\phi_{L,\bm{k}}(\mu_f)\rangle
=
\begin{pmatrix} 
0 \\ 1 
\end{pmatrix}, 
$$
the finite-time matrix elements
reduce to
\begin{equation}
|U_{22}|^2=e^{-\pi\lambda^2},
\qquad
|U_{12}|^2=\frac{1-e^{-\pi\lambda^2}}{\gamma},
\end{equation}
 which yields 
\begin{equation}
\label{probability}
p_k
=
\frac{|U_{22}|^2}{|U_{12}|^2+|U_{22}|^2}
=
\frac{\gamma}{e^{\pi\lambda^2}+\gamma-1}.
\end{equation}
As expected, in the Hermitian limit $\gamma=1$, Eq.~(\ref{probability})
reduces to the  well-known Landau-Zener transition probability,
\begin{equation}
p_k=e^{-\pi\lambda^2}.
\end{equation}
These excitation probabilities are used in the numerical evaluation of the
biorthogonal return rate and the DTOP in the following section.

\section{Numerical Results}

In this section, we present the numerical results obtained from the analytical
solution developed above. We consider a linear ramp of the chemical potential,
$\mu(t)=\mu_f+vt$, beginning sufficiently deep in the left gapped normal
phase, where the system is prepared in its instantaneous ground state. Unless
otherwise stated, we take $\mu_i=-3$ and vary the final value $\mu_f>\mu_i$.
Depending on $\mu_f$ and $\gamma$, the ramp crosses either one or two critical
or exceptional points.

\subsection{Ramp Across a Single Critical Point}

As discussed above, DQPTs can occur only when the ramp ends in a region where
the post-ramp Hamiltonian has a real quasiparticle spectrum. Therefore, for a
ramp crossing a single critical point, we focus on $\gamma>0$. In contrast,
for $\gamma<0$, crossing a single exceptional point drives the system into the
broken-symmetry phase with complex eigenvalues, and DQPTs do not occur.
For $\gamma>0$, the excitation probability depends strongly on the momentum
mode as the ramp crosses the critical point
$\mu_c=-1$ at $k=\pi$. At the gap-closing momentum $k=\pi$, the off-diagonal
pairing terms in the Hamiltonian [Eq.~(\ref{eq:LZ-Hamiltonian})] vanish, so the evolution is
completely nonadiabatic and the excitation probability reaches its maximum,
$p_{k=\pi}=1$. In contrast, away from the critical momentum the quasiparticle
gap remains finite, allowing the system to evolve adiabatically. Consequently,
the excitation probability decreases continuously and approaches
$p_k\rightarrow0$ as $k\rightarrow0$.
The coexistence of these two limiting behaviors,
$p_{k=\pi}=1$ and $p_k\rightarrow0$ for $k\rightarrow0$, together with the
continuity of $p_k$ in the thermodynamic limit, guarantees the existence of a
critical momentum $k^*$ satisfying
$p_{k^*}=1/2$. This critical mode gives rise to DQPTs.

The transition probability is shown as a function of momentum in
Figs.~\ref{fig2}(a)–\ref{fig2}(c) for sweep velocities
$v=0.5$, $v=1$, and $v=2$, respectively, and for several values of the
non-Hermiticity parameter $\gamma$. Throughout this subsection, the chemical
potential is ramped from $\mu_i=-3$ to $\mu_f=0.5$.
Since the ramp crosses a single critical point at $\mu_c=-1$, the excitation
probability satisfies $p_{k=\pi}=1$ and becomes negligible away from the
gap-closing mode ($k\rightarrow0$). Consequently, a unique critical momentum
$k^*$ satisfying $p_{k^*}=1/2$ always exists. This critical mode generates a
sequence of critical times $t_n^*$ according to Eq.~(\ref{eq6}).

The sweep velocity $v$ modifies the excitation probability and therefore shifts
the location of the critical momentum $k^*$. As a result, the sequence of DQPT
critical times changes continuously with the ramp velocity. In particular,
DQPTs persist even in the sudden-quench limit ($v\rightarrow\infty$), where a
critical momentum $k^*$ remains present after crossing a single quantum
critical point.
The degree of non-Hermiticity $\gamma$ has a qualitatively different effect.
Rather than suppressing DQPTs or generating additional critical modes, it
continuously shifts the position of the existing critical momentum $k^*$.
Only at the singular point $\gamma=0$ does the behavior change qualitatively,
where the critical mode disappears and the DQPT structure is modified.

%
%%%%%%%%%%%%%%%%%%%%%%%  Fig.3   %%%%%%%%%%%%%%%%%%%%%%%
\begin{figure*}
\begin{minipage}{\linewidth}
\centerline{\includegraphics[width=0.33\linewidth]{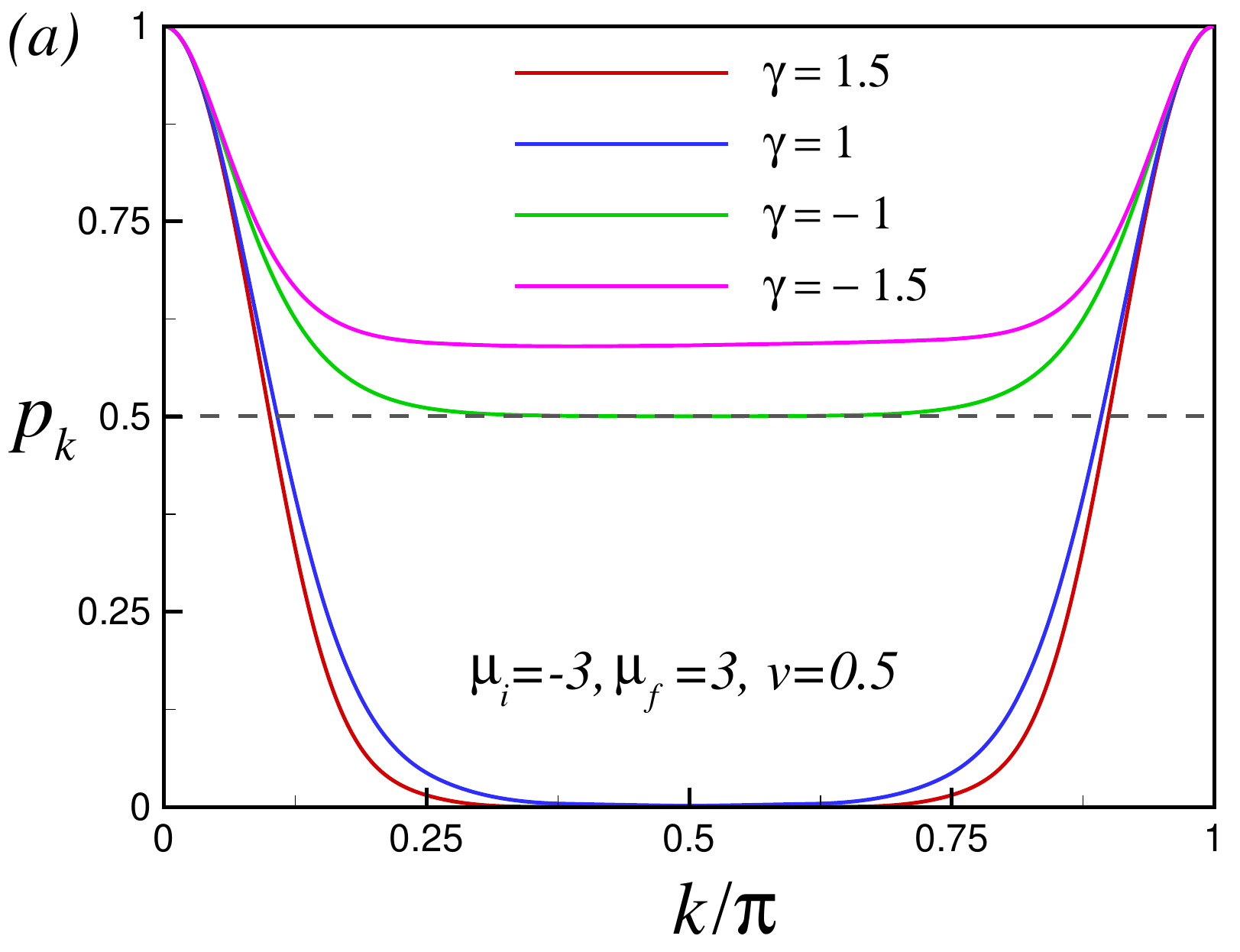}
\includegraphics[width=0.33\linewidth]{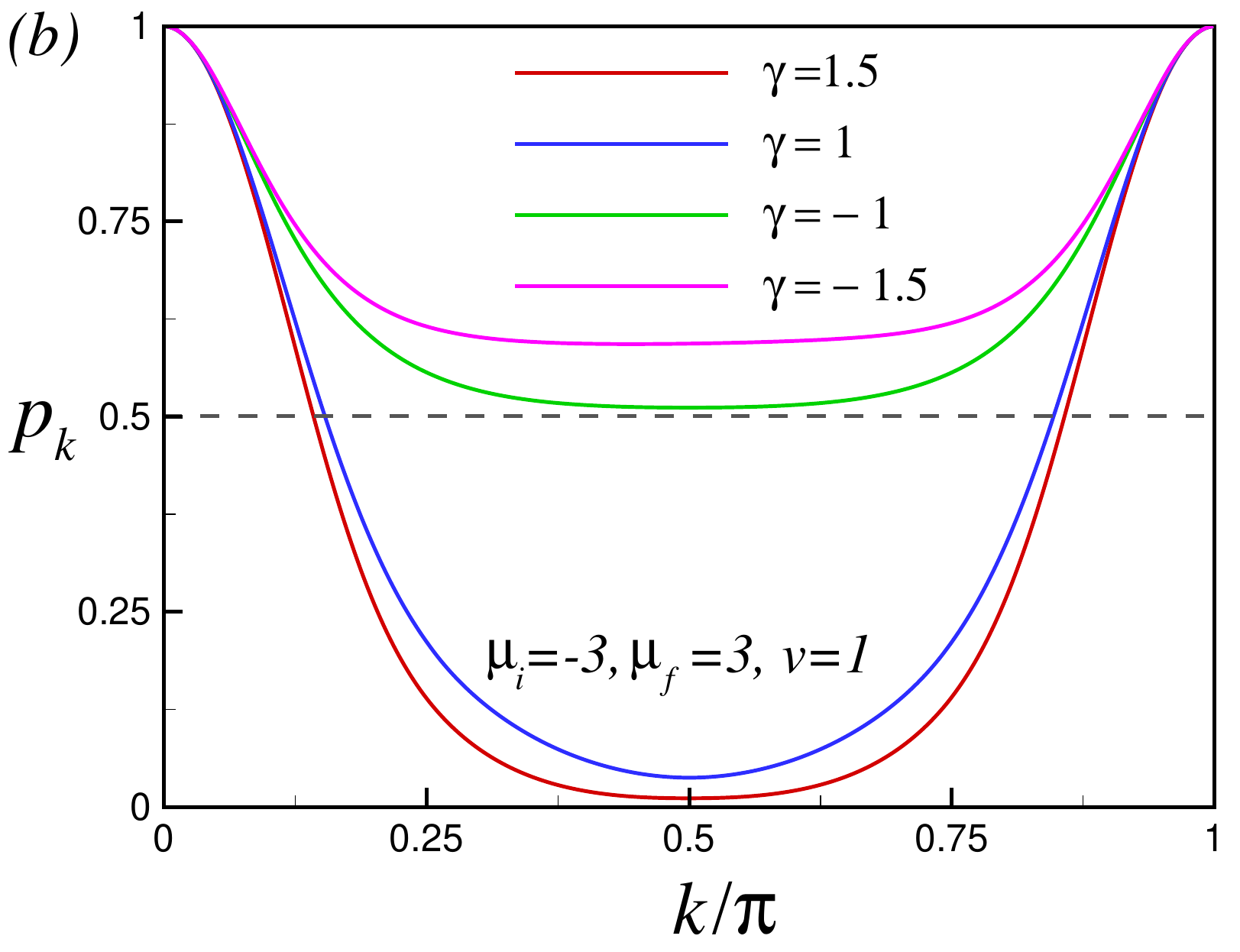}
\includegraphics[width=0.33\linewidth]{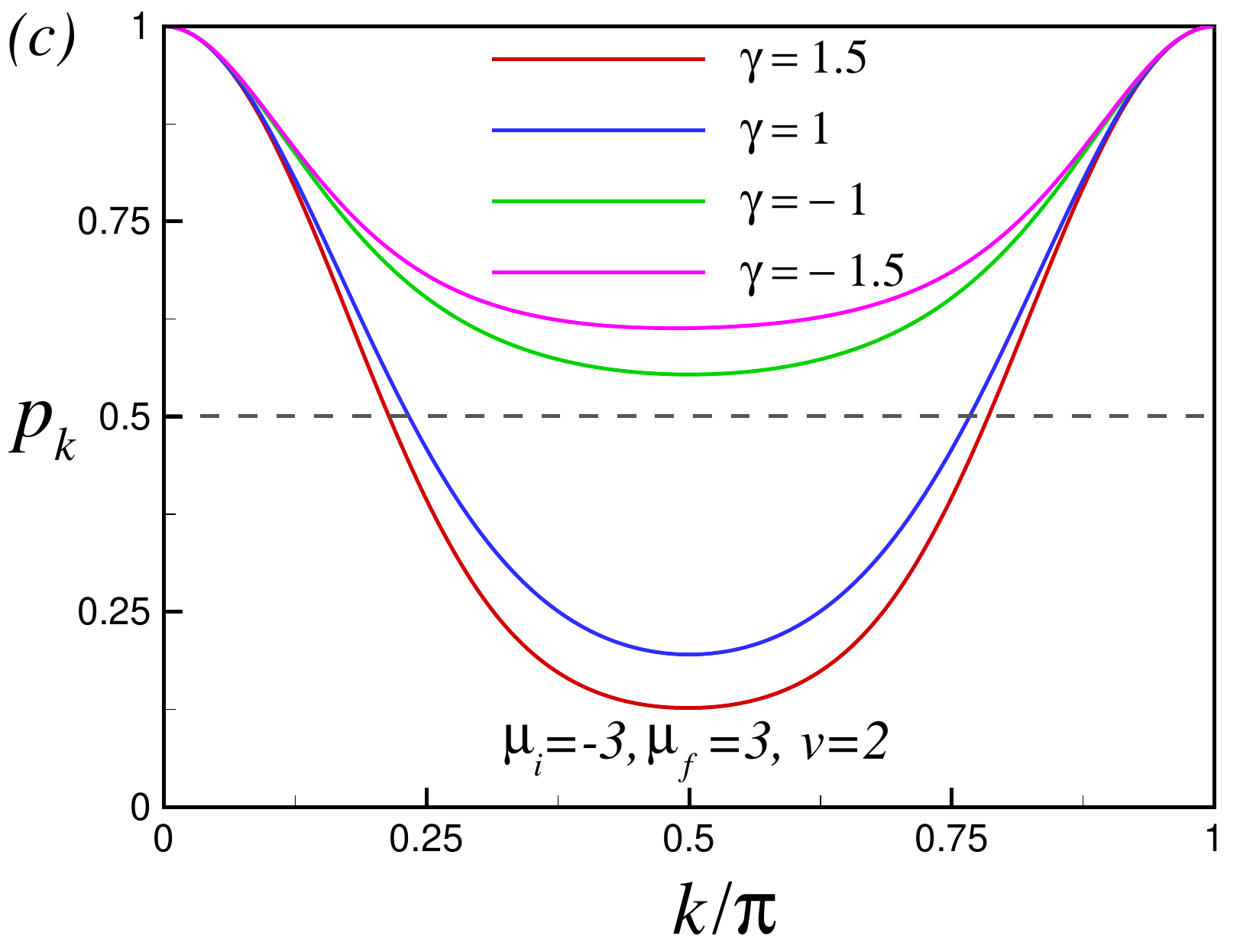}}
\centering
\end{minipage}
%===============================================================
\begin{minipage}{\linewidth}
\centering
\centerline{
\includegraphics[width=0.31\linewidth]{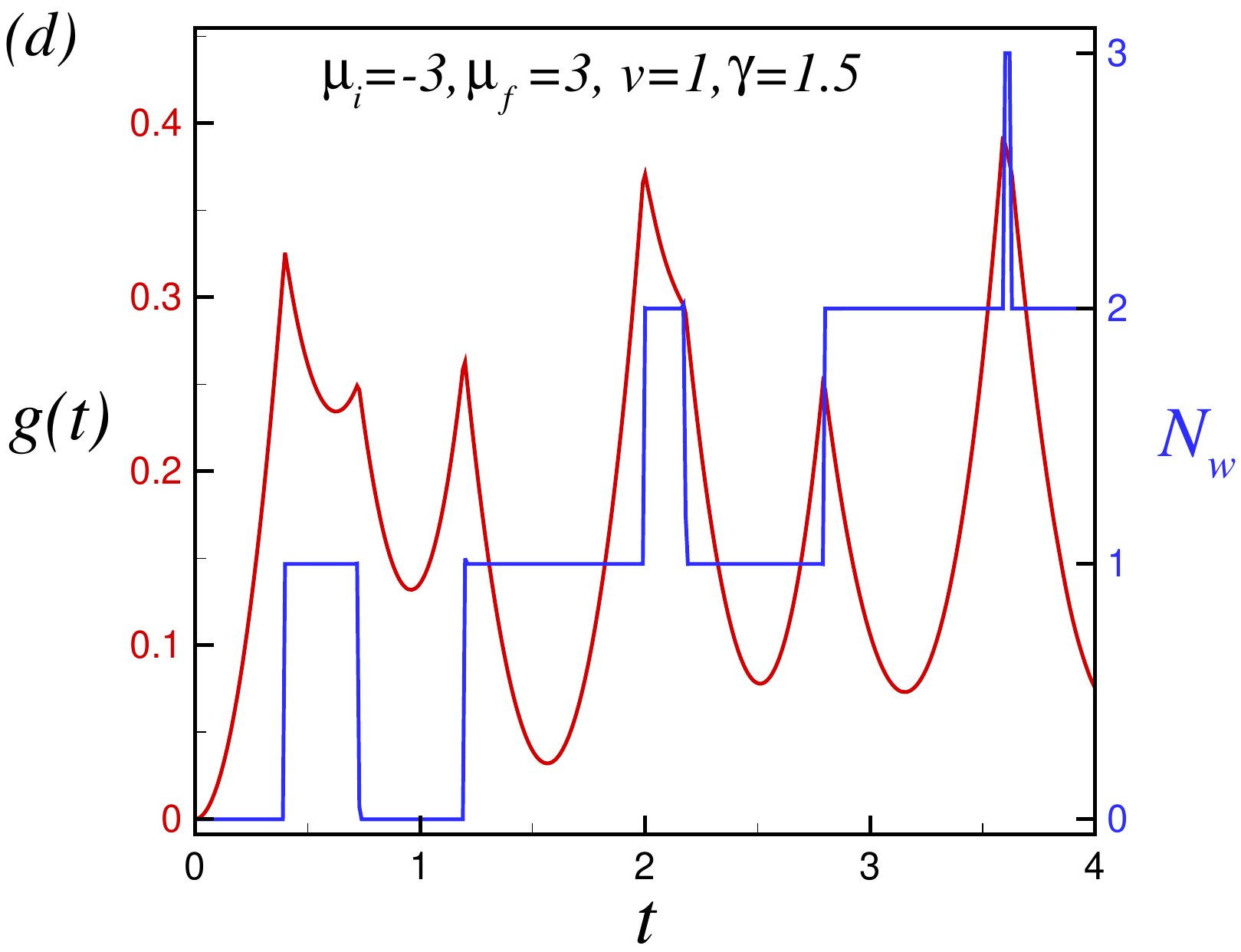}
\hfill
\includegraphics[width=0.31\linewidth]{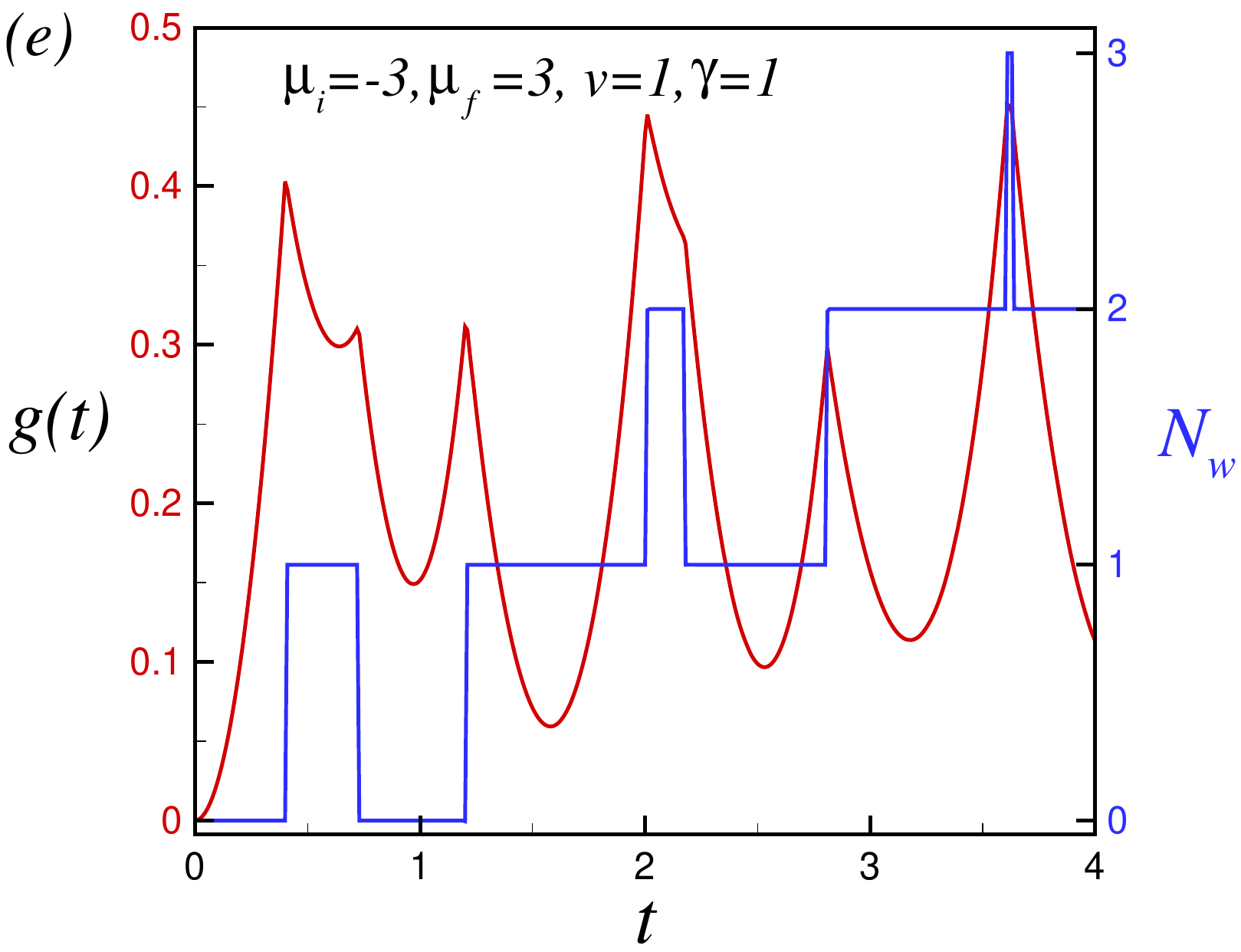}
\hfill
\includegraphics[width=0.31\linewidth]{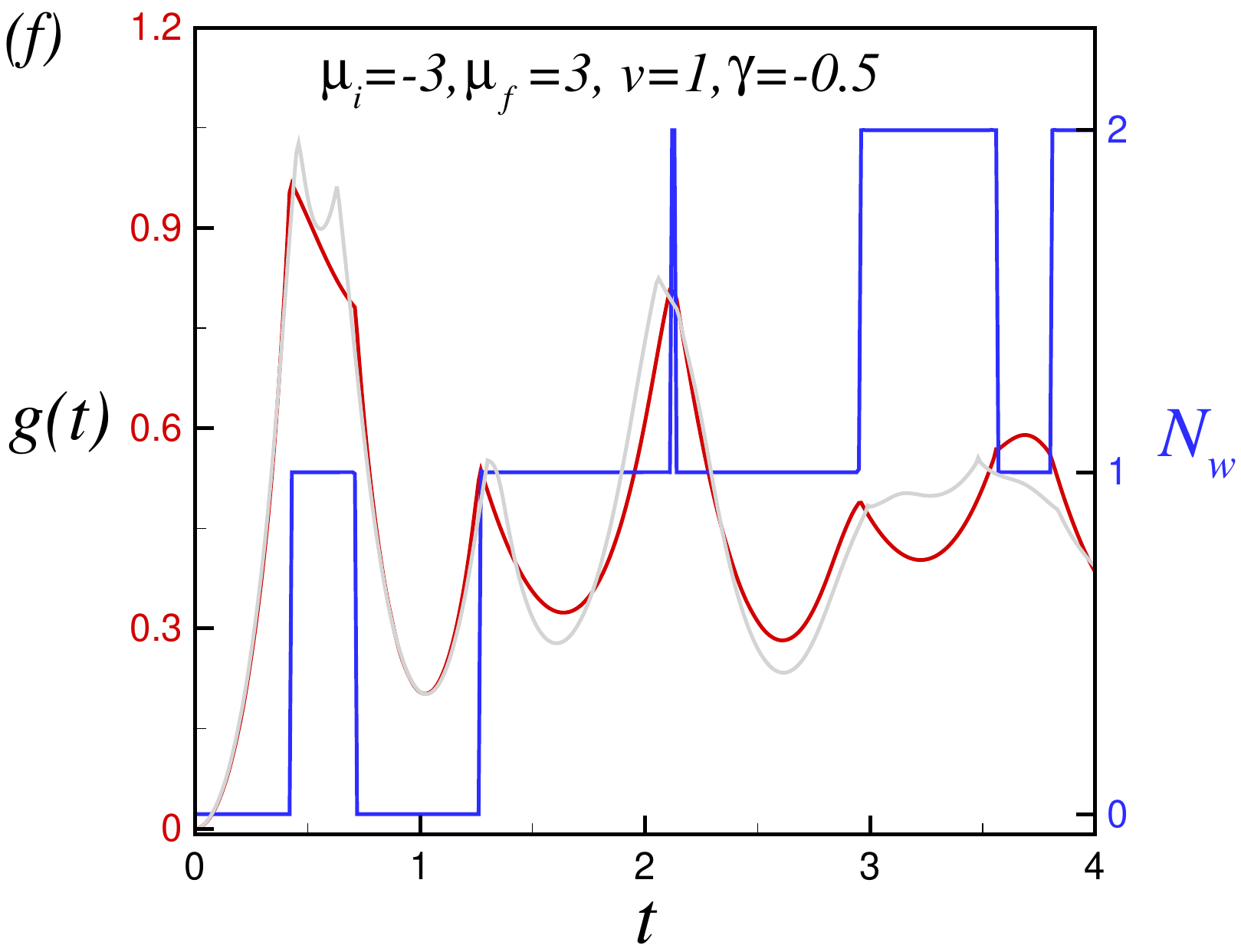}
}
\centering
\end{minipage}
%===============================================================
\caption{Transition probability $p_k$ as a function of momentum for ramps from
$\mu_i=-3$ to $\mu_f=3$, crossing two critical points, $\mu_c=\pm1$, or two
exceptional points, $\mu_{\rm ex}=\pm\sqrt{w^2-\gamma\Delta^2}$. Panels
(a)--(c) show $p_k$ for different values of $\gamma$ and sweep velocities:
(a) $v=0.5$, (b) $v=1$, and (c) $v=2$. Panels (d)--(f) show the corresponding
biorthogonal dynamical free energy $g(t)$ and DTOP $N_w(t)$ for $v=1$:
(d) $\gamma=1.5$, (e) $\gamma=1$, and (f) $\gamma=-0.5$. The blurred curve in
panel (f) denotes the conventional dynamical free energy, shown for comparison.
The system size is $N=1000$.}
\label{fig3}
\end{figure*}
%===============================================================

The dynamical free energy $g(t)$ and the dynamical topological order parameter, $N_w(t)$, are shown in Figs.~\ref{fig2}(d)–\ref{fig2}(f) for the ramp
across a single critical point (corresponding to the excitation probabilities
in Fig.~\ref{fig2}(b)) with a sweep velocity $v=1$ and for
$\gamma=0.1$, $1$, and $1.5$, respectively.
For all three values of $\gamma$, the dynamics are governed by a single
critical momentum $k^*$, which generates one family of critical times
$t_n^*$. At these times, the dynamical free-energy density develops
nonanalytic cusps, signaling DQPTs, while the DTOP changes by one unit and
provides a topological signature of each transition. The direction of the
DTOP jump is determined by the slope of the excitation probability $p_k$ at
the critical momentum $k^*$.
A positive slope leads to a unit increase of
$N_w(t)$, whereas a negative slope results in a unit decrease
\cite{Divakaran2016,Dutta2017}.

Successive jumps (or consecutive drops)
 of the DTOP indicate that the dynamics
are governed by a single critical momentum, whereas the coexistence of both
jumps and drops signals the presence of multiple critical modes. As shown in
Figs.~\ref{fig2}(d)–\ref{fig2}(f), the uninterrupted sequence of jumps confirms
that the ramp across a single critical point is governed by a unique critical
mode.
It is worth noting that the blurred curve in Fig.~\ref{fig2}(d) shows the
conventional dynamical free energy, plotted for comparison with the
biorthogonal dynamical free energy. The critical times obtained from the two
constructions do not coincide: those extracted from the biorthogonal return
rate occur systematically earlier than those obtained from the conventional
one.

%----------------------------------------------
\subsection{Ramp Across Two Critical Points}

We now consider ramps that cross both critical points, $\mu_c=\pm1$, for
$\gamma>0$, or both exceptional points,
$\mu_{\rm ex}=\pm\sqrt{1-\gamma}$, for $\gamma<0$. This protocol reveals a
different dynamical behavior from the single-critical-point ramp. In this case,
the chemical potential is swept from one gapped normal phase to another. 
For a sudden quench across two critical points, DQPTs are generally not
expected~\cite{Vajna2014,Divakaran2016,Zamani2024,Baghran2024,Ansari2025,Kheiri2025}.
In contrast, for a ramp crossing both critical points, the dynamics are
controlled by the gap-closing modes $k=0$ and $k=\pi$, where the
off-diagonal terms in the Hamiltonian vanish.
Therefore, the
dynamics are frozen at these modes, yielding
$p_{k=0}=p_{k=\pi}=1$. In contrast, the minimum of $p_k$ occurs near the mode
with the largest gap, $k=\pi/2$, which is farthest from the gap-closing
momenta. Since $p_{k=0,\pi}=1>1/2$, the occurrence of DQPTs requires the
minimum of the excitation probability to fall below $1/2$. Thus, a sufficiently
slow ramp, $v<v_c$, produces two critical momenta $k_\alpha^*$ and
$k_\beta^*$ satisfying $p_{k_\alpha^*}=p_{k_\beta^*}=1/2$, which generate a
sequence of DQPTs.

Figures~\ref{fig3}(a)--\ref{fig3}(c) show the transition probability as a
function of momentum for a ramp from $\mu_i=-3$ to $\mu_f=3$, crossing two
critical points, $\mu_c=\pm1$, for $\gamma>0$, or two exceptional points,
$\mu_{\rm ex}=\pm\sqrt{1-\gamma}$, for $\gamma<0$. Different panels correspond
to different sweep velocities.
As expected, $p_{k=0}=p_{k=\pi}=1$. For sufficiently slow ramps and
$\gamma>-1$, the minimum of $p_k$ away from the gap-closing modes becomes
smaller than $1/2$. In this case, two critical momenta,
$k_\alpha^*$ and $k_\beta^*$, satisfy
$p_{k_\alpha^*}=p_{k_\beta^*}=1/2$, producing two families of DQPT critical
times, $t_{n,\alpha}^*$ and $t_{n,\beta}^*$, with $n=0,1,\ldots$. Increasing
the sweep velocity raises the minimum of $p_k$. Above a critical sweep velocity
$v_c$, the condition $p_k=1/2$ is no longer satisfied, and DQPTs disappear.
The critical sweep velocity decreases as the degree of non-Hermiticity increases,
corresponding here to reducing $\gamma$ toward the staggered-pairing limit
$\gamma=-1$.

The dynamical free energy and the associated DTOP for ramps across two
critical or exceptional points are shown in Figs.~\ref{fig3}(d)--\ref{fig3}(f)
for $v=1$ and different values of $\gamma$, corresponding to the excitation
probabilities in Fig.~\ref{fig3}(b). The DQPTs appear as nonanalytic cusps in
$g(t)$ and, more clearly, as steps in $N_w(t)$. The nonsequential jumps and
drops of the DTOP indicate the presence of two critical momenta with opposite
slopes of $p_k$. For comparison, the blurred curve in Fig.~\ref{fig3}(f)
shows the conventional dynamical free energy plotted together with the
biorthogonal return rate.

%
%%%%%%%%%%%%%%%%%%%%%%%  Fig.4   %%%%%%%%%%%%%%%%%%%%%%%
\begin{figure}[t]
\centerline{\includegraphics[width=\linewidth]{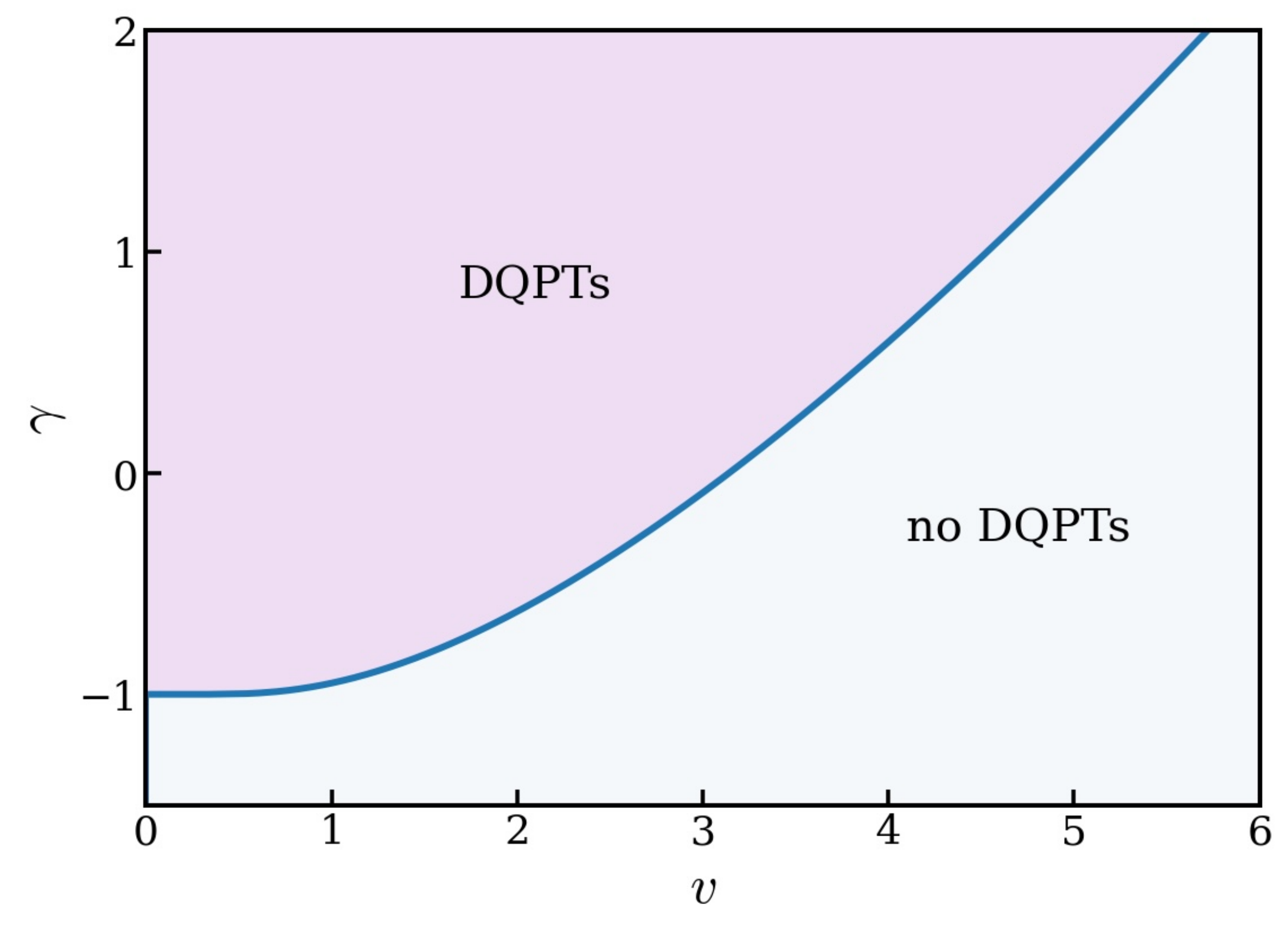}}
\caption{Dynamical phase diagram in the $v$-$\gamma$ plane for a ramp across
two critical points or two exceptional points. The region marked ``DQPTs''
indicates parameter values for which dynamical quantum phase transitions occur.}
\label{fig4}
\end{figure}
%%%%%%%%%%%%%%%%%%%%%%%%%%%%%%%%%%%%%%%%%%%%%%%%%%%%%%%
%

As discussed above, DQPTs occur only when the minimum of $p_k$ becomes smaller
than $1/2$, which is satisfied for $v\leq v_c$. Equation~(\ref{probability})
gives the excitation probability for the ramp from $\mu_i\to-\infty$ to
$\mu_f\to+\infty$. Since the minimum of $p_k$ occurs at $k=\pi/2$, the critical
sweep velocity is obtained from the condition $p_{k=\pi/2}=1/2$, yielding
\begin{equation}
v_c=\frac{\pi\gamma\Delta^2}{\ln(1+\gamma)}.
\end{equation}
This expression is valid for $\gamma>-1$, with the value at $\gamma=0$
defined by continuity as $v_c=\pi\Delta^2$. In the Hermitian limit,
$\gamma=1$, it reduces to $v_c=\pi\Delta^2/\ln 2$, while
$v_c\to0$ as $\gamma\to-1^{+}$.
The dynamical phase diagram in the $v$-$\gamma$ plane is shown in
Fig.~\ref{fig4} for a ramp from $\mu_i\to-\infty$ to $\mu_f\to+\infty$. The
region marked ``DQPTs'' supports sequences of dynamical quantum phase
transitions, while DQPTs are absent outside this region. 
As shown, the critical sweep velocity  decreases as the system is driven deeper into the
non-Hermitian regime, i.e., as $\gamma$ is reduced toward $-1$.

%%%%%%%%%%%%%%%%%%%%%%%%%%%%%%%%%%%%%%%%%%%%%%%%%%%%%%%%%%%%%%%%%%%%%%%%%%%%%%
\section{Summary and Conclusion}
%%%%%%%%%%%%%%%%%%%%%%%%%%%%%%%%%%%%%%%%%%%%%%%%%%%%%%%%%%%%%%%%%%%%%%%%%%%%%%

We have investigated the nonequilibrium dynamics of the
imbalanced-pairing Kitaev model, a representative pseudo-Hermitian
superconducting system, subjected to a linear ramp of the chemical
potential. Using the biorthogonal formulation of quantum mechanics,
we developed a general framework for describing dynamical quantum
phase transitions (DQPTs) in non-Hermitian superconductors with
particle-number nonconserving pairing interactions.
The independent momentum sectors of the model were mapped onto an
effective non-Hermitian Landau-Zener problem, allowing the
time-dependent Schr\"odinger equation to be solved analytically.
This yielded exact expressions for the excitation probabilities,
which determine the biorthogonal Loschmidt echo, the dynamical
free-energy density, and the associated dynamical topological order
parameter.

Our analysis demonstrates that DQPTs require the post-ramp Hamiltonian to
possess a real quasiparticle spectrum, together with the existence of a
critical momentum $k^*$ satisfying $p_{k^*}=1/2$. In the
time-reversal-symmetric pseudo-Hermitian regime, $\gamma>0$, a ramp
across a single critical point always generates a unique critical
momentum satisfying $p_{k^*}=1/2$, giving rise to a single family
of critical times. The degree of non-Hermiticity continuously shifts
the location of the critical momentum but neither suppresses the
DQPTs nor generates additional critical modes.
For ramps crossing two critical points, the dynamics are governed by
two critical momenta provided that the minimum excitation probability
falls below one half. Consequently, DQPTs survive only for sweep
velocities below a critical value $v_c$, which we obtained
analytically. We showed that $v_c$ decreases as $\gamma$ is reduced
toward the staggered-pairing limit $\gamma=-1$, where it vanishes.
For $\gamma<-1$, DQPTs disappear entirely.

The accompanying DTOP provides a topological characterization of the
nonequilibrium dynamics. A single critical momentum leads to
successive jumps of the winding number, whereas two critical
momenta produce alternating jumps and drops, reflecting the topology
of the excitation probability in momentum space.
Our results establish a direct connection between
pseudo-Hermiticity, exceptional-point physics, and dynamical quantum
phase transitions under continuous driving. Beyond the present
model, the analytical framework developed here can be applied to other integrable non-Hermitian systems, providing a useful platform for exploring
nonequilibrium topology and critical dynamics in experimentally
realizable open and engineered quantum systems.

%%%%%%%%%%%%%%%%%%%%%%%%%%%%%%%%%%%%%%
%%%%%%%%%%%%%%%%% Acknowledgments %%%%%%%%%%%%%%%%%%%%%
%	
\section*{Acknowledgments}
A. A. acknowledges financial support from the Beijing Natural Science Foundation under Grant No. IS25015. 
The work of H. Y. was supported by the Beijing Natural Science Foundation under Grant No. IS23013.
 The authors thank
S. Mahdavifar for useful comments and discussions.

%%%%%%%%%%%%%%%
%%%%%%%%%%%%%%%
%%%%%%%%%%%%%%%
\appendix
%%%%%%%%%%%%%%%
%%%%%%%%%%%%%%%
%%%%%%%%%%%%%%%%%%%%%%%%%%%%%%%%%%%%%%%%%%%%%%%%%%%%%%%%%%%%%%%%%%%%%%%%

%%%%%%%%%%%%%%%%%%%%%%%%%%%%%%%%%%%%%%%%%%%%%%%%%%%%%%%%%%%%%%%%%%%%%%%%%%%%%%

%%%%%%%%%%%%%%%%%%%%%%%%%%%%%%%%%%%%%%%%%%%%%%%%%%%%%%%%%%%%%%%%%%%%%%%%%%%%%%
\section{Symmetry Properties and Pseudo-Hermiticity}
\label{app:symmetry}
%%%%%%%%%%%%%%%%%%%%%%%%%%%%%%%%%%%%%%%%%%%%%%%%%%%%%%%%%%%%%%%%%%%%%%%%%%%%%%
This appendix summarizes the symmetry properties of the imbalanced-pairing Kitaev model, including pseudo-Hermiticity, time-reversal symmetry, particle-hole symmetry, and the role of exceptional points. These symmetry properties govern the transition between the time-reversal-symmetric phase with a purely real quasiparticle spectrum and the broken-symmetry phase characterized by complex-conjugate eigenvalues.
\paragraph{Time-Reversal Symmetry:}
The imbalanced-pairing Kitaev Hamiltonian is invariant under the antiunitary time-reversal operator $\mathcal{T}$, which acts as $\mathcal{T}i\mathcal{T}^{-1}=-i$ and $\mathcal{T}c_j\mathcal{T}^{-1}=c_j$. Accordingly, $\mathcal{T}\mathcal{H}\mathcal{T}^{-1}=\mathcal{H}$. Unlike conventional Hermitian systems, where time-reversal symmetry is simply compatible with Hermiticity, here it coexists with a non-Hermitian Hamiltonian and is responsible for the characteristic pseudo-Hermitian spectral structure. As long as this symmetry remains unbroken, the quasiparticle spectrum is entirely real. 
\paragraph{Pseudo-Hermiticity:}
For $\gamma \neq 0$, each momentum sector of the Bogoliubov-de Gennes Hamiltonian satisfies the exact pseudo-Hermiticity relation
\begin{equation}
H_k^{\dagger} = \eta H_k \eta^{-1},
\end{equation}
where the metric operator in the Nambu basis is given by
\begin{equation}
\eta=
\begin{pmatrix}
1&0\\ 0&\gamma^{-1}
\end{pmatrix}.
\end{equation}
The nature of the metric dictates the spectral properties of the system:
\begin{itemize}
    \item For $\gamma>0$, the metric $\eta$ is positive definite. The Hamiltonian is quasi-Hermitian and can be mapped to an equivalent Hermitian Hamiltonian, guaranteeing a purely real spectrum.
    \item For $\gamma<0$, the metric $\eta$ is indefinite. Pseudo-Hermiticity alone does not guarantee a real spectrum; it merely implies that the eigenvalues are either real or appear in complex-conjugate pairs. The transition to complex eigenvalues occurs via the spontaneous breaking of time-reversal symmetry.
\end{itemize}
\paragraph{Singularity at $\gamma=0$:}
The point $\gamma=0$ is singular: the metric $\eta$ becomes non-invertible, and pseudo-Hermiticity is strictly undefined. At this point, the Bogoliubov-de Gennes matrix becomes strictly upper triangular. When $\mu_k=w\cos k-\mu = 0$ while $\Delta_k \neq 0$, the matrix becomes a nonzero nilpotent Jordan block, rendering the Hamiltonian defective. Therefore, $\gamma=0$ must be treated separately from both the positive-definite quasi-Hermitian regime and the indefinite pseudo-Hermitian regime.
\paragraph{The $\mathcal{PT}$-Symmetric Limit:}
At $\gamma=1$, the Hamiltonian is strictly Hermitian and also satisfies parity-time ($\mathcal{PT}$) symmetry, where the parity operator exchanges momenta $k \leftrightarrow -k$. For generic $\gamma \neq 1$, the relevant mathematical structure supporting the real spectrum is pseudo-Hermiticity rather than $\mathcal{PT}$ symmetry. Furthermore, the case of $\gamma=-1$ (staggered pairing) possesses fundamentally distinct spectral properties from $\gamma=1$ due to the relative minus sign in the energy gap equation, $\varepsilon_k^2 = \mu_k^2 - \Delta_k^2$, which actively permits the emergence of exceptional points. 

\paragraph{Particle-Hole Symmetry:}
For the Bogoliubov-de Gennes Hamiltonian considered here, the Nambu structure imposes a generalized pseudo-particle-hole symmetry. With $\mathcal{K}$ denoting complex conjugation, the generalized antilinear pseudo-particle-hole operation is $\mathcal{C}=U_C\mathcal{K}$, with the linear part constructed using the metric operator as
\begin{equation}
U_C
=
\eta^{-1}\sigma_x
=
\begin{pmatrix}
0 & 1\\
\gamma & 0
\end{pmatrix}.
\end{equation}
For $\gamma \neq 1$, $U_C$ is not unitary with respect to the ordinary Hilbert inner product. By direct matrix multiplication, one verifies that this linear part satisfies the symmetry condition:
\begin{equation}
U_C H_k^* U_C^{-1} = -H_{-k}.
\end{equation}
This fundamental constraint ensures that the spectrum is symmetric about zero energy: if $E_k$ is an eigenvalue of $H_k$, then $-E_k^*$ is an eigenvalue of $H_{-k}$.
\paragraph{Exceptional Points:}
Exceptional points occur when the gap closes at $(w\cos k-\mu)^2 + \gamma\Delta^2\sin^2k = 0$. Unlike ordinary Hermitian degeneracies, exceptional points are characterized by the simultaneous coalescence of both the quasiparticle energies and their corresponding eigenvectors. At these points, the Hamiltonian becomes defective and the biorthogonal basis %presented in %Appendix~\ref{app:diagonalization} 
breaks down. For $\gamma<0$, these exceptional points serve as the phase boundaries separating the parameter regions with purely real quasiparticle spectra from the regions with complex-conjugate eigenvalue pairs.

%%%%%%%%%%%%%%%%%%%%%%%%%%%%%%%%%%%%%%%%%%%%%%%%%%%%%%%%%%%%%%%%%%%%%%%%%%%%%%
%%%%%%%%%%%%%%%%%%%%%%%%%%%%%%%%%%%%%%%%%%%%%%%%%%%%%%%%%%%%%%%%%%%%%%%%%%%%%%
\section{Exceptional-Point Topology and Point-Gap Remarks}
\label{app:nh-topology}

This appendix summarizes the aspects of non-Hermitian topology that are
relevant to the imbalanced-pairing Kitaev model. We focus on point-gap
structures, exceptional points, and their relation to the dynamical topology
discussed in the main text. The invariant distinguishing the topological and
non-topological superconducting phases in Fig.~\ref{fig1} is the
pseudo-Hermitian continuation of the Kitaev-chain winding or Pfaffian
invariant and has been discussed in Ref.~\cite{Li2018}. Since the present work
focuses on DQPTs, we do not rederive that equilibrium invariant here.
Instead, we emphasize the intrinsically non-Hermitian spectral structure
associated with exceptional points and clarify its distinction from the DTOP.

%%%%%%%%%%%%%%%%%%%%%%%%%%%%%%%%%%%%%%%%%

\paragraph{Point-gap topology and winding number:}
A non-Hermitian two-band Hamiltonian can be written in the form
\begin{equation}
H_k=d_x(k)\sigma_x+i d_y(k)\sigma_y+d_z(k)\sigma_z,
\end{equation}
where $d_x(k)$, $d_y(k)$, and $d_z(k)$ are real functions. Unlike Hermitian systems, where topology is usually defined with respect to a real energy gap, non-Hermitian Hamiltonians may also possess a point gap in the complex-energy plane. In one dimension, such a point gap can be characterized by the winding of a complex function around a reference point.

For the present class of two-band models, one may introduce the complex function
\begin{equation}
z(k)=d_x(k)+i d_y(k).
\end{equation}
For the imbalanced-pairing Kitaev model, the components are defined as $d_x(k) = \frac{1+\gamma}{2}\Delta\sin k$ and $d_y(k) = \frac{1-\gamma}{2}\Delta\sin k$, yielding the complex function:
\begin{equation}
z(k) = \frac{\Delta\sin k}{2} [(1+\gamma) + i(1-\gamma)].
\end{equation}
Because $z(k)$ traces a one-dimensional linear trajectory passing directly through the origin rather than forming a closed loop, the conventional point-gap winding number evaluated at the reference point $z=0$ is ill-defined (since $z(k)=0$ at $k=0$ and $k=\pi$). For reference points away from the line traced by $z(k)$, the winding is simply trivial (zero). 

Most importantly, the point-gap winding defined purely from $z=d_x+id_y$ ignores $d_z(k) = w\cos k - \mu$. Therefore, it cannot distinguish $|\mu|<w$ from $|\mu|>w$, and is not the invariant that defines the topological and non-topological superconducting phase boundary in Fig.~1. The topological features unique to the non-Hermitian aspects of this model are instead dictated by the coalescence of the bands at exceptional points.

\paragraph{Exceptional points and defective Hamiltonians:}
A defining feature of non-Hermitian systems is the possible appearance of exceptional points. At an exceptional point, not only do two eigenvalues become degenerate, but the corresponding eigenvectors also coalesce. The Hamiltonian is then defective and cannot be diagonalized by a complete set of eigenvectors. Locally, it can be brought to a Jordan-block form,
\begin{equation}
H_{\rm EP}\sim\begin{pmatrix}\varepsilon & 1\\ 0 & \varepsilon
\end{pmatrix},
\end{equation}
where $\varepsilon$ is the degenerate eigenvalue. This distinguishes exceptional points from ordinary Hermitian degeneracies, where eigenvectors remain linearly independent.

In the imbalanced-pairing Kitaev model, the exceptional points are determined
by the gap-closing condition in
Eq.~(\ref{exceptional-point-condition}). At these points, the two
quasiparticle eigenvalues and their corresponding eigenvectors coalesce,
rendering the Hamiltonian defective and the biorthogonal basis singular.
In the vicinity of an exceptional point, the radicand of the dispersion,
$
\varepsilon_k
=
\sqrt{(w\cos k-\mu)^2+\gamma\Delta^2\sin^2k},
$
vanishes linearly in the relevant two-dimensional parameter space. As a
result, the quasiparticle energies form a square-root branch point.
Encircling the exceptional point exchanges the two eigenvalue branches and
gives rise to a half-integer non-Hermitian topological charge.

For $\gamma<0$, this condition separates parameter regions with purely real quasiparticle spectra from regions with complex-conjugate eigenvalue pairs. These exceptional points therefore form the boundary between the time-reversal-symmetric and broken-symmetry phases discussed in the main text.

\paragraph{Relation to the dynamical topology:}
The static topological features discussed above characterize the equilibrium band structure. In contrast, the main text characterizes DQPTs using the dynamical topological order parameter $N_w(t)$, which is constructed from the gauge-invariant Pancharatnam geometric phase of the biorthogonal Loschmidt amplitude. These two notions of topology are conceptually related through the winding of complex phases, but they play different roles in the present work. The static exceptional-point topology provides background for the non-Hermitian band structure, whereas the DTOP directly diagnoses the real-time dynamical transitions.

%%%%%%%%%%%%%%%%%%%%%%%%%%%%%%%
\bibliography{NHKC_DQPTs}
%%%%%%%%%%%%%%%%%%%%%%%%%%%%%%%

%%%%%%%%%%%%%%%%%%%%%%%%%%%%%%%
\end{document}